# Glassy Carbon Microelectrode Arrays Enable Voltage-Peak Separated Simultaneous Detection of Dopamine and Serotonin Using Fast Scan Cyclic Voltammetry

*Elisa Castagnola[a, e†], Sanitta Thongpang[b,c,e], Mieko Hirabayashi[a,e], Giorgio Nava[d], Surabhi Nimbalkar[a,e], Tri Nguyen[a,e], Sandra Lara [a,e], Alexis Oyawale[e], James Bunnell[e], Chet Moritz[b,e], Sam Kassegne[a,e]\**

[a] NanoFAB. SDSU Lab, Department of Mechanical Engineering, San Diego State University, San Diego, CA, USA

[b] University of Washington, Departments of Electrical & Computer Engineering, Rehabilitation Medicine, and Physiology & Biophysics, Seattle, WA

[c] Department of Biomedical Engineering, Faculty of Engineering, Mahidol University, Nakorn Pathom, Thailand

[d] Department of Mechanical Engineering, University of California, Riverside, Riverside, CA, USA,

[e] Center for Neurotechnology (CNT), Bill & Melinda Gates Center for Computer Science & Engineering; Seattle, WA 98195, USA

\* corresponding author: Sam Kassegne • Professor of Mechanical Engineering, NanoFAB.SDSU Lab, Department of Mechanical Engineering, College of Engineering, San Diego State University, 5500 Campanile Drive, CA 92182-1323. E-mail: kassegne@sdsu.edu • Tel: (760) 402-7162.

KEYWORDS Glassy Carbon, Neural Probes, Microelectrode Arrays, C-MEMS, Fast Scan Cyclic Voltammetry, *in vivo* Neurotransmitter Detection, Dopamine, Serotonin, Simultaneous Detection

**ABSTRACT**

Progress in real-time, simultaneous *in vivo* detection of multiple neurotransmitters will help accelerate advances in neuroscience research. The need for development of probes capable of stable electrochemical detection of rapid neurotransmitter fluctuations with high sensitivity and selectivity and sub-second temporal resolution has, therefore, become compelling. Additionally, a higher spatial resolution multi-channel capability is required to capture the complex neurotransmission dynamics across different brain regions. These research needs have inspired the introduction of glassy carbon (GC) microelectrode arrays on flexible polymer substrates through carbon MEMS (C-MEMS) microfabrication process followed by a novel pattern transfer technique. These implantable GC microelectrodes offer unique advantages in electrochemical detection of electroactive neurotransmitters through the presence of active carboxyl, carbonyl, and hydroxyl functional groups. In addition, they offer fast electron transfer kinetics, capacitive electrochemical behavior, and wide electrochemical window. Here, we combine the use of these GC microelectrodes with the fast scan cyclic voltammetry (FSCV) technique to optimize the co-detection of dopamine and serotonin *in vitro* and *in vivo*. We demonstrate that using optimized FSCV triangular waveform at scan rates ≤ 700 V/s and holding and switching at potentials of 0.4 and 1V respectively, it is possible to discriminate voltage reduction and oxidation peaks of serotonin and dopamine, with serotonin contributing distinct multiple oxidation peaks. Taken together, our results present a compelling case for a carbon-based MEA platform rich with active functional groups that allows for repeatable and stable detection of electroactive multiple neurotransmitters at concentrations as low as 10 nM.

1. Introduction

Innovative neural probes are becoming increasingly critical for both uncovering fundamental principles in neuroscience and providing therapeutic intervention in a variety of neurological disorders (Jorgenson et al. 2015; Lerner et al. 2016; Moritz 2018). Recent progress in clinical neuromodulation and brain computer interfaces (BCIs) have been enabled by substantial progress in signal recording and stimulation hardware to continuously monitor the nervous system subsequently deliver appropriate stimulation for closed-loop control (Chapman et al. 2018; Lee et al. 2017; Nimbalkar et al. 2018; Parastarfeizabadi and Kouzani 2017). Further, the development of implantable multi-modal probes capable of reading and writing not only electrophysiological but also electrochemical neural signals (Alcami and Pereda 2019; Miller et al. 2015; Pereda 2014) may become a key enabler of understanding brain function. This integration of neurochemical reading with electrophysiology recording and stimulation is of fundamental importance in elucidating the relationship between electrical and electrochemical signaling and their role in the pathogenesis of neurological disorders (Alcami and Pereda 2019; Miller et al. 2015; Pereda 2014). This increased understanding will also inform novel treatments, and, therefore, in improving their treatments (Bozorgzadeh et al. 2015; Chang et al. 2013; Grahn et al. 2014; Lee et al. 2017), such as neurochemical feedback in smart adaptive deep brain stimulation (DBS) systems (Chang et al. 2013; Grahn et al. 2014; Lee et al. 2017).

A variety of electrochemical techniques have been used to monitor neurotransmitter levels in vivo (Crespi et al. 1988; Ferapontova 2017; Gratton et al. 1988; Njagi et al. 2010; Sharma et al. 2018b). Among these, fast scan cyclic voltammetry (FSCV) is preferred due to its sub-second scale high temporal resolution (hundreds of millisecond range) that is consistent with scale of chemical fluctuations at neuronal synapsis (Hashemi 2013; Oh et al. 2016; Ou et al. 2019; Robinson et al. 2003; Swamy and Venton 2007). For the past 30 years, FSCV has been commonly used in combination with carbon fiber electrodes (CFEs) that exhibit excellent spatial resolution with minimal tissue damage and inflammatory response due to their small size (7-10 μm in diameter) (Castagnola et al. 2020; Ou et al. 2019; Puthongkham and Venton 2020). However, these CFEs often lack in selectivity and experience signal degradation over a period of time due to biofouling (Harreither et al. 2016; Hensley et al. 2018; Puthongkham and Venton 2020). They are also typically limited to a single-site recording (Dankoski and Wightman 2013; Puthongkham and Venton 2020; Swamy and Venton 2007), even though there have been several reported attempts to improve their spatial resolution by fabricating CFE arrays that have shown promising electrochemical detection and physiological recordings (Patel et al. 2016; Schwerdt et al. 2017a; Schwerdt et al. 2017b). Their fabrication, however, consists of a time-consuming manual process that do not allow for facile batch-fabrication high-density arrays. On the other hand, several strategies have been adopted to both improve the selectivity of specific neurotransmitters and decrease electrode biofouling. These include the functionalization of CFEs with charged polymers (Peairs et al. 2011; Raju et al. 2019; Taylor et al. 2017; Vreeland et al. 2015), size exclusion membranes (Qi et al. 2016; Zhou et al. 2019), or even sp3-hybridized carbon materials (Puthongkham et al. 2018; Roberts and Sombers 2018; Swamy and Venton 2007; Zestos et al. 2015), or FSCV waveform optimization (Hensley et al. 2018; Jackson et al. 1995; Puthongkham and Venton 2020). However, despite some promising improvements in selectivity and anti-fouling properties, real-time simultaneous detection of multiple neurotransmitter concentrations *in vivo* using FSCV remains a challenge, in particular for dopamine (DA) and serotonin (5-hydroxytryptamine, 5-HT) - two of the key electrochemical analytes in the central nervous system. This task is further complicated by the fact that, with fast scan rates, both DA and 5-HT have similar oxidation potentials that makes distinguishing them difficult (Puthongkham and Venton 2020; Swamy and Venton 2007).

5-HT and DA are involved in the regulation of primary neural functions and play a critical role in the regulation of behavior. For example, 5-HT regulates mood and sleep functions and is a major target for pharmacological treatment of depression (Celada et al. 2013; Fakhoury 2016; Schloss and Williams 1998; Yohn et al. 2017). DA, on the other hand, is involved in motor control, reward, motivation and cognitive function (Felger and Treadway 2017; Haber 2010; Klein et al. 2019; Wise 2004). Additionally, their interactions can be implicated in the development and progression of neuropsychiatric disease, such as Parkinson's Disease (Boileau et al. 2008; Carta et al. 2008; Politis et al. 2012; Wong et al. 1995), schizophrenia (Kapur and Remington 1996; Niederkofler et al. 2015), depression (Boileau et al. 2008; Dremencov et al. 2004; Zangen et al. 2001), and Tourette syndrome (Steeves and Fox 2008; Wong et al. 2008). Thus, the simultaneous detection of DA and 5-HT in vivo in real time and the study of their interactions stands out as an area of significant research interest. However, except the early works of Swamy et al. who reported simultaneous detection in vivo using CNT-modified CFEs (Swamy and Venton 2007) and Zhou et al.(Zhou et al. 2005) who performed in striatal slices using bare CFEs, the literature in simultaneous detection of these neurotransmitters in vivo is very sparse. In the case of in vitro experiments, simultaneous detection has been reported using PEI-CNT fibers (Zestos et al. 2014). In these FSCV studies, DA and 5-HT presented an oxidation peak at the same potential, but they can be differentiated by their reduction peak. Other in vitro studies have been reported using mm-size modified electrodes and slower temporal resolution voltammetric techniques (Han et al. 2014; Selvaraju and Ramaraj 2003; Selvaraju and Ramaraj 2005; Wang et al. 2003; Wu et al. 2003; Zhang et al. 2006). For example, *Wang et al.* have demonstrated that electrodes made of intercalated CNT on graphite surface can simultaneously detect DA and 5-HT, in the presence of AA (Wang et al. 2003). They attributed the high sensitivity and selectivity to the unique carbon surface of the nanotubes and the porous interfacial layers due to the subtle tubule structure of CNT. Kachoosangi and Compton used planar edge-plane pyrolytic electrodes polished with alumina powder to simultaneously detect DA, 5-HT and ascorbic acid (AA) *in vitro* (Kachoosangi and Compton 2007). They reported well-resolved voltammetric peaks for the direct oxidation of DA, 5-HT and AA in phosphate buffer, thanks to the use of edge plane pyrolytic graphite. The authors emphasized that, to the best of their knowledge, all prior literature reported in detection of DA and 5-HT in the presence of AA had used modified electrodes.

Therefore, to address these challenges and allow the integration of multi-site neurochemical detection into multimodal closed-loop systems, progress is required in developing (a) materials rich with electrochemically-active functional groups and good adsorption characteristics, (b) electrochemical measurement protocols optimized for improved sensitivity and selectivity in an environment of complex kinetics of neuronal chemicals on microelectrode surfaces, and (c) microfabrication techniques that yield array of implantable carbon-based microelectrode arrays (MEAs) for multisite measurements.

From the perspective of the need for developing new materials and processes to respond to the need of multi-site measurement, we recently introduced a pattern transfer technology for the integration of glassy carbon (GC) microelectrodes, pre-microfabricated on silicon wafer through a high-temperature carbon-MEMS process, with flexible polymer substrates (Castagnola et al. 2018; Nimbalkar et al. 2018). An advantage of this process is the potential to batch-fabricate implantable GC MEAs in a highly reproducible way, opening significant opportunities for a wider use of GC microelectrodes in neural hardware applications. GC has subsequently emerged as a compelling material for microelectrodes of neural probes, as it offers competitive sensitivity and selectivity like CFEs, with the added advantage of tunable mechanical and electronic properties enabled by its range of possible hybridized bonds ($sp^2$ and $sp^3$) together with chemical inertness, biocompatibility, good electrical properties, electrochemical stability, purely capacitive charge injection, and fast surface electrochemical kinetics (Nimbalkar et al. 2018; Puthongkham and Venton 2020). Indeed, our previous works have demonstrated that GC MEAs are capable of high-quality electrophysiological recording and stimulation with outstanding electrochemical stability and biocompatibility (Castagnola et al. 2018; Goshi et al. 2018; Nimbalkar et al. 2018; Vahidi et al. 2020; Vomero et al. 2017). These GC MEAs also have higher sensitivity compared to CFEs due to the presence of numerous basal planes rich in functional groups[4]. Thus, the GC MEAs open up the possibility for integrated multimodal electrophysiological and electrochemical measurement on the same arrays (Castagnola et al. 2018; Nimbalkar et al. 2018).

A fundamental understanding of the mechanisms driving adsorption of electroactive species, such as DA and 5-HT, on GC microelectrode surfaces will help develop optimized detection protocols. Here, we investigate the electrochemical kinetics of DA and 5-HT at planar GC microelectrodes using a variety of FSCV waveforms to optimize co-detection of DA and 5-HT *in vitro* and *in vivo*. To enable a better understanding of adsorption/desorption kinetics of DA, 5-HT and their combination, we also investigate the use of multi-waveform FSCV (*M*-FSCV), a powerful technique that provides additional information on adsorption/desorption characteristics of neurotransmitters (Kim et al. 2018).

## 2. Materials and Methods

### 2.1 Microfabrication

We microfabricated a 4-channel penetrating neural probes on flexible polymeric substrate with a total shank length of 7 mm (and 0.5 mm width) for targeting the rat striatum and four GC microelectrode detection sites (1500 $\mu m^2$ area), positioned in the striatum, with an inter-electrode distance of 220 μm as shown in Figure 1 a.

The core extended C-MEMS microfabrication technology used for the fabrication of the GC microelectrode arrays supported on polymeric substrates is described in detail elsewhere (Castagnola et al. 2018; Goshi et al. 2018; Vomero et al. 2017). This recently introduced technique consists of a pattern transfer method that enabled the incorporation of pre-patterned GC microelectrodes on flexible polyimide substrate, expanding the use of GC technology to implantable neural probes suited for electrophysiological and electrochemical recordings and electrical stimulation (Nimbalkar et al. 2018). Here, we further extend the functionality of this microfabrication technology by adding a reinforcing layer to allow easy penetration of brain tissue, in order to target deep brain regions (Castagnola et al. 2018; Vahidi et al. 2020).

In summary, the microfabrication process involves spin-coating SU8 negative photoresist (Microchem, MA) at 1200 rpm for 55 s and soft-baking at 65°C for 10 min and 95°C for 20 min followed by UV exposure at ~400 mJ/cm$^2$. The post-exposure bake consists of 65°C for 1 min and 95 °C for 5 min. This was followed by development of SU8 for 3–5 min and curing at 150°C for 30 min. Pyrolysis was done at 1000°C in an inert $N_2$ environment following protocols described elsewhere (Vomero et al. 2017; Vomero et al. 2016), resulting in GC microelectrodes with high graphitic content (Hirabayashi et al. 2013). Briefly, pyrolysis is carried out in a closed quartz tube-furnace under vacuum and Nitrogen atmosphere through gradual heating to 1000 °C followed by cooling to room temperature (Vomero et al. 2016). After the pyrolysis step, 6 μm layer of photo-patternable polyimide (HD 4100) (HD Microsystems, DE, USA) was spin-coated on top of GC microelectrodes at 2500 rpms for 45 s, soft baked at 90°C for 3 min and at 120 °C for 3 min, then cooled down to room temperature, and patterned through UV exposure at ~400 mJ/cm$^2$. Post-exposure bake consisted of 80°C for 1 min. Development was performed using a spray-puddle process where QZ3501 (Fuji Film, Japan) was dispersed to form a puddle on a stationary wafer. A rinse was applied after a set time of 15 s, followed by spin-drying of the wafer (2000 rpm for 15 s and 500 rpm s$^{-1}$ ramp). The spray-puddle cycle was repeated three times and the wafer rinsed with SU8 developer (MicroChem, USA). Subsequently, the polyimide layer was partially cured at 300°C for 60 min under a $N_2$ environment.

Following, metal traces were deposited using NR91000PY negative photoresist (Futurrex Inc., USA) as a sacrificial layer. NR91000PY was spin-coated at 500 rpm for 45 s and ramped down for 10 s, then prebaked for 2 min at 150 °C followed by 380 mJ/cm$^2$ UV exposure. Post exposure bake was done at 100 °C for 2 min and the sample was developed in RD6 developer (Futurrex Inc., USA) for 3 s. Subsequently, 20 nm Ti adhesion layer and 200 nm Pt layer were deposited through sputtering. After metal deposition, a lift-off process was performed, and the sacrificial layer was removed in acetone. For electrical insulation, an additional 6 μm of polyimide HD4100 (300 rpms) was spun, patterned (400 mJ/cm$^2$), and cured (350°C for 90 min) under $N_2$ environment. Additional 30 μm thicker layer of polyimide (Durimide 7520, Fuji Film, Japan) was spin-coated (800 rpm, 45 s) and then patterned (400 mJ/cm$^2$) on top of the insulation layer to reinforce the penetrating portion of the device. Then it was developed, as previously described, and final cured at 350 °C for 90 min. Subsequently, the device was released from the wafer through selective etching of silicon dioxide with buffered hydrofluoric acid. The probes were then connected to a custom-built printed circuit board (PCB) that served as the connector to the FSCV system (Figure 1 a).

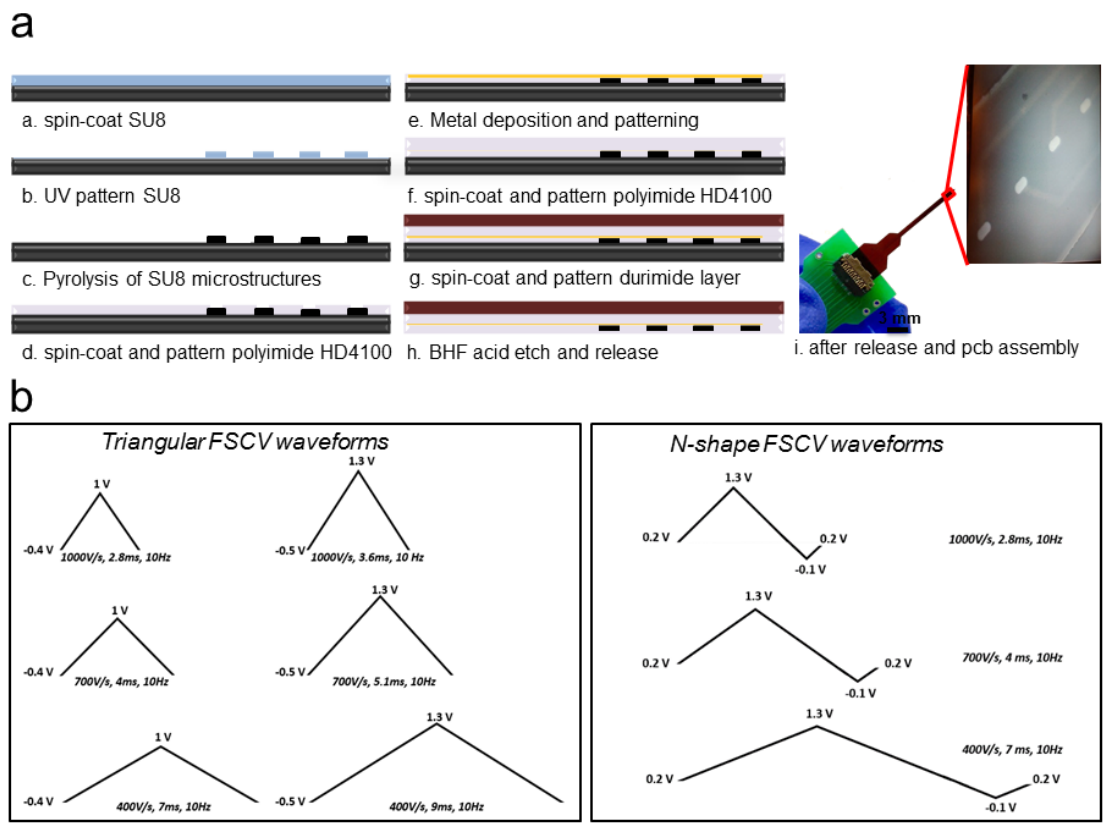

**Figure 1** (a) microfabrication steps (left) and a 4-channel penetrating neural probes on polymeric substrate (right), with a total shank length of 7 mm (and 0.5 mm width) for targeting the rat striatum and four GC microelectrode detection sites (1500 μm$^2$ area), positioned in the striatum, with an inter-electrode distance of 220 μm (inset). The probe was connected to a custom-built printed circuit board (PCB) that served as the connector to the FSCV system. (b) FSCV electrochemical waveforms used for the DA and 5-HT detection and co-detection: triangular FSCV with EW -0.4/1V and -0.5/1.3V at 400,700 and 1000V/s, respectively (left) and N-shaped FSCV (0.2 to 1.3 to −0.1 to 0.2 V) at 400, 400,700 and 1000V/s, respectively (right).

**2.2 Simultaneous Detection and Microelectrode Kinetics Experiments**

FSCV was performed using Wave Neuro Potentiostat System (Pine Research, NC). As shown in Figure 1 b, we used FSCV waveforms consisting of (a) triangular FSCV waveforms at three different scan rates (400, 700, and 1000 V/s) and two sets of holding (i.e., -0.4 and -0.5V) and switching potentials (1 and 1.3 V). The corresponding electrochemical windows (EW) were -0.4V/1V and -0.5V/1.3V, respectively, and (b) *N*-shaped modified FSCV waveform at 3 different scan rate (400,700 and 1000 V/s) and holding and switching potentials of (0.2 to 1.3 and −0.1 to 0.2 V). The *N*-shaped waveform used here is a modified version of the Jackson waveform (Jackson et al. 1995). This waveform was designed to reduce fouling reactions of 5-HT's oxidative and reductive by-products at CFEs, improving electrode sensitivity and stability over time (Jackson et al. 1995; Puthongkham and Venton 2020).

Prior to the beginning of each experiment, the same voltage waveform was applied to the microelectrodes at 60 Hz for 15-20 minutes for activating the carbon surface of the microelectrodes (Pavel Takmakov 2010). For electrode calibration, known concentrations of DA, 5-HT, and their mixture were then infused over 5 seconds while changes in current were recorded for 20 seconds. For the kinetics experiments, known concentration of DA or 5-HT were injected into the PBS solution and then changes in current were recorded for 60 seconds. For the co-detection experiments, the same concentration of their mixture (50% DA: 50% 5-HT) was simultaneously added to the PBS solution and then changes in current were recorded for 60 seconds.

### 2.3 *In Vivo* Experiments

**Acute FSCV Experiments in Rat Brain:** All animal experiments were performed in accordance with the Association for Assessment and Accreditation of Laboratory Animal Care (AAALAC) Guide for the Care and Use of Laboratory Animals (8th Edition) and approved by the University of Washington Institutional Animal Care and Use Committee (IACUC) under protocol number 4265-01. Adult female Long-Evans rats (250-300g) were used in this study and anesthesia was induced with Urethane (1.5 g kg-1, i.p., made in a 50% w/w solution in 0.9% saline). The animal was placed in a stereotaxic frame and the GC probe targeting caudate-putamen (relative to bregma: AP +1.2, ML +2.0, DV -4.5) and Ag/AgCl reference electrode were placed contralateral of the recording electrode. Additional hole was drilled above the substantia nigra area at AP -5.6, ML +1.4, DV -8.0 for a stimulating wire electrode. The dorsoventral position of the stimulating electrode was adjusted until peak and robust stimulated release was obtained.

**Voltammetry Recording Sessions:** As described in Section 2.3, a triangle waveform was applied to the GC microelectrode with ramping from - 0.4 V to 1 V and back (vs. Ag/AgCl reference) at a rate of 400 V/s and frequency of 10 Hz. This waveform enabled the discrimination of oxidation and reduction DA and 5-HT peaks *in vitro*. Stimulation train of 60 pulses was applied at 60 Hz with 2 ms width per pulse at 250 µA. For all studies, stimulations were performed every 3 minutes. After five baseline stimulations were recorded, carbidopa (25 mg kg-1 in 0.9% saline, i.p., Sigma Aldrich) was administered to block peripheral decarboxylases and thirty minutes later, 5-hydroxytyptophan (5-HTP, Sigma Aldrich) was administered (200 mg kg-1 in 0.9% saline, s.c.) [22]. was administered (200mg kg-1 in 0.9% saline, s.c.) (Swamy and Venton 2007).

### 2.4 Materials Characterization

TEM imaging of the synthesized GC electrodes was carried out on a Tecnai12 microscope. The carbonaceous material was removed from the substrate with a blade, dispersed into chloroform and drop-casted on a copper TEM grid (the solvent was evaporated at room temperature). The Raman Spectra of the synthesized electrode materials were recorded in the spectral range 800–3900 cm$^{-1}$ using a micro Raman Horiba LabRam microscope (laser wavelength 532 nm, laser power 0.06 mW, 50× objective).

## 3 Results and Discussions

### 3.1 *In Vitro* FSCV Characterizations

In this section, we present the outcomes of the *in vitro* electrochemical sensing performance of the GC microelectrodes for FSCV detection of DA, 5-HT and their mixture. We focus on the evaluation of the adsorption kinetics of DA and 5-HT at the GC surface as a function of scan rate, holding and switching potentials (i.e. electrochemical windows), and holding potential time, in order to obtain a better DA and 5-HT peak discrimination.

For *in vitro* electrochemical kinetics experiments, we used low concentrations of DA and 5-HT (10 nM - 200 nM). This selection was guided by the previously reported high sensitivity of similarly prepared GC microelectrodes (Castagnola et al. 2018; Nimbalkar et al. 2018).

#### 3.1.1 Separate Detection of Dopamine and Serotonin

First, experiments on separate detection of DA and 5-HT are presented to help understand the adsorption kinetics of these two neurochemicals at GC microelectrodes. Particular focus is placed on identifying their oxidation and reduction peaks using different FSCV waveforms and how these peaks are influenced by scan rates and voltage sweep ranges, i.e. holding and switching potentials. This will guide the adoption of the most appropriate FSCV waveform that will result in separate and distinct peaks corresponding to DA and 5-HT.

DA is an electroactive neurotransmitter that electrochemically oxidizes in a two-electron oxidation described by the equation: DA → DOQ + 2e- + 2H+, where DOQ is the *o-quinone* form of DA (Bath et al. 2000; Kim et al. 2018) (Scheme 1).

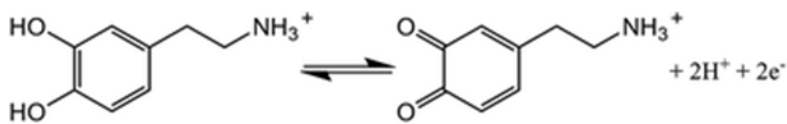

Scheme 1: two-electron, two-proton oxidation of dopamine. Adapted from(Chen et al. 2018).

DA is typically detected with FSCV through a waveform commonly called "*the dopamine waveform*", where a holding potential of −0.4 V is applied to the working electrode to selectively preconcentrate cationic DA on the electrode surface (Roberts and Sombers 2018; Rodeberg et al. 2016; Venton and Wightman 2003). Then, a triangular waveform with a scan rate of 400 V/s is applied at 10 Hz to scan the electrode to a switching potential of +1/+1.3V and back to -0.4V to oxidize dopamine and reduce *dopamine-o-quinone* (Puthongkham and Venton 2020; Roberts and Sombers 2018; Venton and Wightman 2003). The 10 Hz frequency guarantees 100 ms temporal resolution, sufficient for capturing rapid neurotransmitter release in the brain (Puthongkham and Venton 2020; Venton and Wightman 2003).

As shown in Figure 2 a, FSCV using triangular waveforms (10 Hz) for DA shows that the oxidation and reduction kinetics for DA at GC microelectrodes seems to slow down with an increase in scan rates from 400 – 1000 V/s. This may be due to insufficient time available for completion of DA oxidation under fast scan rates in the range of thousands of V/s (Venton and Cao 2020; Wightman and Wipf 1990). The separation between reduction and oxidation peaks (ΔE) was observed to increase with higher scan rates, shifting the oxidation peaks to the right and the reduction peaks to the left, respectively (Figure 2 a, Supplementary Figure 2a). For EW of -0.4/1V, the oxidation peaks corresponding to 400 and 700 V/s scan rates increased from 0.65±0.05V, to 0.78±0.03V (Table 1), while the reduction peaks decreased from -0.22±0.03V to -0.30±0.05V, respectively, resulting in ca 17% increase in ΔE from 0.87±0.05 V to 1.04±0.05V. However, using a scan rate of 1000V/s in the same EW, the DA kinetics was too slow to allow a proper discrimination of the oxidation and reduction peaks (Figure 2 a). Further, ΔE increase is observed for all DA concentrations, EW, and FSCV waveforms used (Supplementary Figure 2 a-c, Supplementary Figure 3 d-f).

Similar DA peak shifts at higher scan rates and higher EW voltages have been observed for CFEs. Again, this is due to the insufficient time available for completion of DA oxidation under fast scan rates in the range of thousands of V/s. During FSCV, the electron transfer kinetics for DA at surface of carbon electrodes is slow (Deakin et al. 1986; Venton and Cao 2020) and the shape of the voltammogram is influenced by scan rate, rate of electron transfer, and current density (Venton and Cao 2020). Further, the DA adsorption/desorption seemed to show slower kinetics at GC electrodes with respect to the results reported for CFEs in the literature (Bath et al. 2000; Venton and Cao 2020). This is probably due to the effect of microelectrode size and form factor, i.e. planar disk microelectrodes (GC) vis-à-vis cylinders (CFEs), that can influence mass transport and time-dependence of the diffusion field (Forster 1994). Ionic current response due to oxidation/reduction of the analyte contains both time-dependent (transient) and a time-independent terms, while ionic current response over a relatively long periods of times is dominated by a time-dependent radial diffusion and is affected by the size and the shape of microelectrodes (Forster 1994). While the exposed areas of the GC microelectrodes are comparable to those of typical CFEs with 7 μm diameter and 70 μm length (in the range of CFEs commonly used in FSCV studies (Cryan and Ross 2019; Hensley et al. 2018; Mitch Taylor et al. 2012), the smaller radius of the CFEs, ca. 3.5 µm is likely to allow for a more rapid steady-state response to changes in the applied potential, as opposed to the slower diffusion in the larger GC microelectrodes.

Furthermore, with scan rate of 400 V/s, a distinct shift in both oxidation and reduction peaks was observed between FSCV at -0.4V/1V (0.65±0.05V, -0.22±0.03V, ΔE = 0.87±0.05) and the wider EW of -0.5V/1.3V (0.79±0.01V, -0.35±0.01V, ΔE = 1.14±0.06V), informing that peaks are also functions of the EW (Supplementary Figure 2 b, Supplementary Figure 3 d-f). Due to the slow DA kinetics at GC electrodes under high scan rates (1000V/s), the wider EW (-0.5/1.3) allowed for the discrimination of the oxidation peak (1.01±0.02V) which was not possible using the -0.4/1V EW (Supplementary Figure 2 b-d). However, this EW was not sufficiently expanded to detect the reduction peak of DA at 0.5V (Supplementary Figure 2 a, Supplementary Figure 3 f). Fouling test in the presence of DA was performed using the triangular waveform at scan rate of 400V/s. The current peak amplitudes in response to 50 nM of DA were stable over the entire recording session. Additionally, no significant drifting was observed in the FSCV background during FSCV recording over a period of 25 minutes, demonstrating the electrochemical stability of the GC surfaces (Supplementary Figure 4).

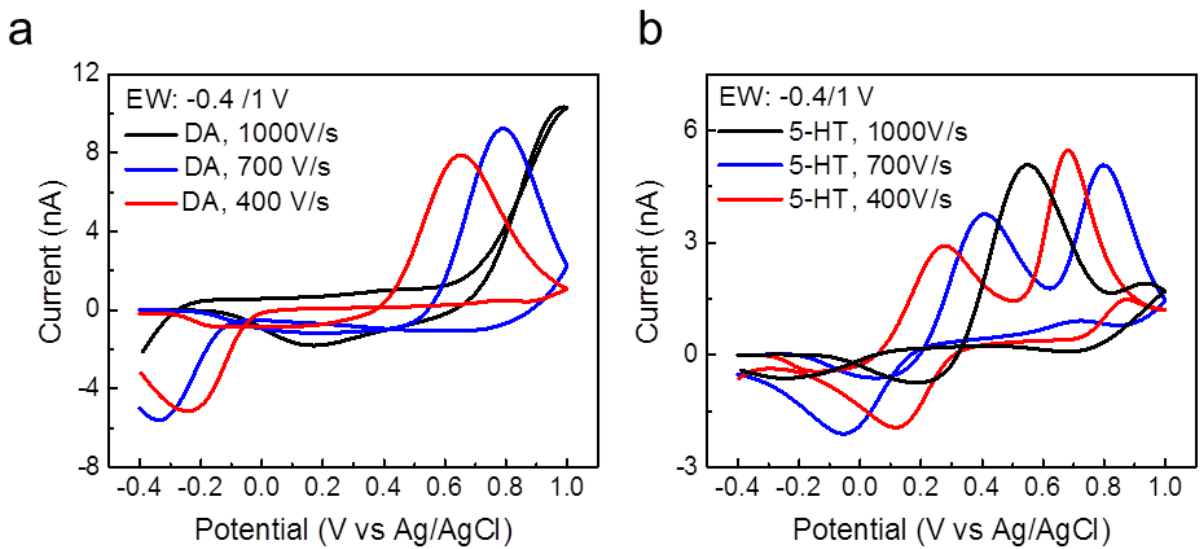

**Figure 2**. Effect of scan rate on DA and 5-HT kinetics using -1/0.4 V EW. (a) Effect of scan rate. DA concentration at 10 nM, 1000V/s, oxidation peak (Ox) > 1V, reduction peak (Redx) < -0.4V (black line); 700 V/s, Ox = 0.78±0.03V, Redx = -0.30 ±0.05V (blue line) ; 400 V/s, Ox = 0.65±0.05V, Redx = -0.22±0.03V, (red line),(b) Effect of scan rate on 5-HT (10 nM) oxidation peaks. While two separate oxidation peaks are observed at lower scan rates (≤ 700 V/s), these peaks merge for scan rate of 1000 V/s. The CV plots correspond to the average of 5 repetitions on 3 different electrodes (see Table 1 and Table 2).

| DA | Oxidation Peak (nA) | Redox Peak (nA) | ΔE (V) |
|---|---|---|---|
| 700V/s small EW | 8.86±0.74-- | -5.47±0.45 | 1.04±0.05 |
| 400V/s small EW | 8.28±1.46 | -4.87±0.84 | 0.87±0.05 |
| 1000V/s small EW | -- | -- | -- |

Table 1: Mean and Standard Deviation (N=3, 5 repetitions each) of the amplitudes of 10 nM dopamine oxidation and reduction peaks and the corresponding peak separation (corresponding to Figure 2 a)

5-HT, just like DA, is an electroactive neurotransmitter that can be electrochemically oxidized within the physiological pH solvent window (Scheme 2) (Wrona and Dryhurst 1990). Its oxidation reaction mechanism involves a multi-step two-electron, two-proton transfer process (Jackson et al. 1995; Patel et al. 2013; Verbiese-Genard et al. 1984), during which by-products such as reactive carbocation intermediate and dimers are formed (Scheme 2) (Jackson et al. 1995; Patel et al. 2013; Wrona and Dryhurst 1990). The Electro-oxidation mechanism of 5-HT has been extensively studied by Wrona *et. al* under acidic (Wrona and Dryhurst 1987, 1990) and physiological conditions (Wrona and Dryhurst 1990). They proposed that oxidation happens in a series of steps where 5-HT is oxidized first to its carbocation, followed by a further oxidation to *aquinone imine* as shown in Scheme 2 (Wrona and Dryhurst 1987, 1990).

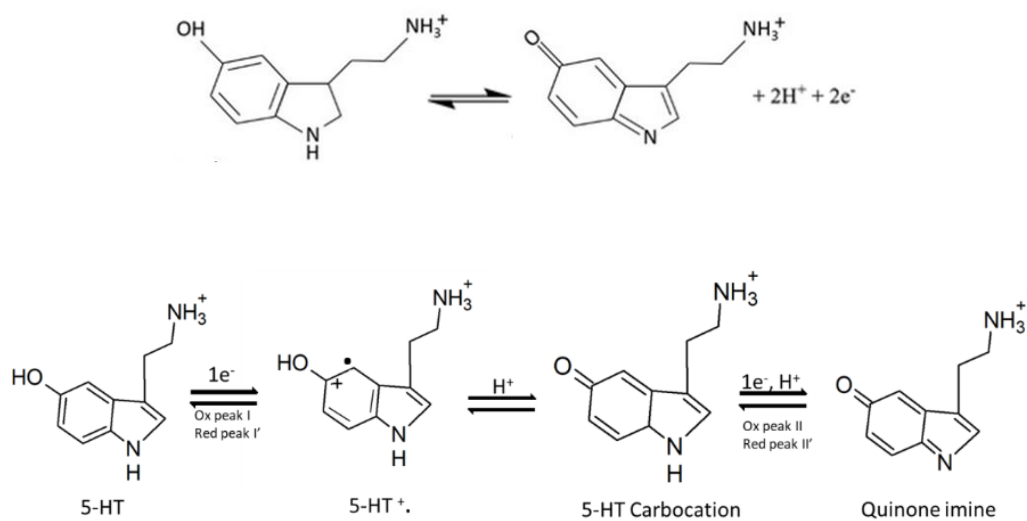

Scheme 2. Mechanism of serotonin oxidation on carbon electrode surfaces adapted from(Mendoza et al. 2020) (top); and two step oxidations proposed and experimentally validated by Wrona *et al.* (Wrona and Dryhurst 1987, 1990) (bottom).

Since this reaction dominates at fast scan rates (Wrona and Dryhurst 1990), more recent literature have shown that this is the primary reaction that occurs at scan rates of 300 - 1000V/s (Güell et al. 2014; Lama et al. 2012; Patel et al. 2013). However, subsequent reactions of the carbocation with 5-HT produce dimers (5,5'-Dihydroxy-4,4'-bitryptamine, 3-(2-Aminoethyl)-3-[3'-(2-aminoethyl)-indol-5-one-4'-yl]-5-hydroxyindolenine, and 5-[[3-(2-Aminoethyl)-1H-indol-4-yl]oxy]-3-(2-aminoeythyl)- H-indole)(Wrona and Dryhurst 1990). The by-products have been shown to be very reactive and adsorb irreversibly on the electrode surface resulting in fouling (Jackson et al. 1995; Patel et al. 2013). To overcome this difficulty, the *in vitro* and *in vivo* detections of 5-HT at CFEs are usually performed using *N*-shaped FSCV waveform (Hashemi et al. 2009; Jackson et al. 1995), also known as Jackson waveform (Jackson et al. 1995), that has been optimized to accelerate electrode response times and reduce the formation of strongly adsorptive by-products. Specifically, this waveform holds the potential at +0.2 V to limit 5-HT by-product adsorption, scan quickly at 1000 V/s to 1.0 V to limit fouling, and switch down to −0.1 V to allow the detection of the reduction peak (Jackson et al. 1995). However, this N-shaped FSCV waveform cannot be efficiently used to detect DA which needs a more negative holding potential to facilitate the cationic adsorption on the electrode surface (Supplementary Figure 3 a).

| 5-HT | Peak-I (nA) | Peak-II (nA) | Redox Peak (nA) | ΔE (Peak-I) (V) | ΔE (Peak-II) (V) |
|---|---|---|---|---|---|
| 700V/s small EW | 3.38±1.91 | 5.82±2.02 | -2.18±0.74 | 0.42±0.02 | 0.82±0.01 |
| 400V/s small EW | 2.90±1.38 | 5.46±1.34 | -1.93±0.33 | 0.17±0.01 | 0.59±0.01 |
| 1000V/s small EW | -- | 5.86±0.91 | -1.00±0.19 | -- | 0.51±0.02 |

Table 2: Mean and Standard Deviation (N=3, 5 repetitions each) of the amplitudes of 10 nM serotonin oxidations (Peak I and II) and reduction peaks and the peak separations (corresponding to Figure 2 b).

As the response of 5-HT is complex and involves multi-reaction steps, the background subtracted FSCV in these experiments exhibited unique double oxidation peaks at scan rates ≤ 700 V/s. For example, as shown in Figure 2 b, for FSCV of 5-HT with EW of -0.4V/1V and 10 nM concentration, two separate oxidation peaks were observed, i.e., 0.27±0.04 (Peak-I) and 0.68±0.03V (Peak-II) for 400 V/s and 0.41±0.01V (Peak-I) and 0.79±0.02V (Peak-II) for 700 V/s scan rates. Using 0.4/1V EW at higher scan rate (1000 V/s), the two oxidation peaks seem to converge to a single peak at 0.53±0.05V (Figure 2 b). This suggests that higher scan rates increase 5-HT kinetics at this EW. At slower scan rates, an overall slower kinetic enables the detection of the peak I (Table 2) that is possibly due to the electro polymerization of the secondary oxidation products of the multi-step 5-HT oxidation reaction (Yang et al. 2017). Using a similar waveform, the same peak at ca. 0.3–0.4 V for 5-HT has been observed at PEI/CNT fibers (Yang et al. 2017). This is potentially problematic since the reactive oxidation products (i.e. reactive carbocation intermediate and dimers) of this secondary reaction of 5-HT (Jackson et al. 1995; Patel et al. 2013; Wrona and Dryhurst 1990) have been demonstrated to form an insulating layer on the surface of the CFEs and decrease sensitivity over time (Jackson et al. 1995; Patel et al. 2013). To address this concern, we investigated the stability of 5-HT electrochemical detection at GC microelectrode surface, where GC microelectrodes were continuously scanned in presence of 50 nM of 5-HT and the 5-HT detection was monitored for 8 h using the triangular FSCV at 400 V/s. Every 20-40 minutes, the PBS solution containing 5-HT was changed followed by the collection of a new background measurement and a new injection of the same 5-HT concentration. The current peak amplitudes in response to 50 nM of 5-HT were stable with no significant drop in detection sensitivity (one-way Anova, $p>0.05$) (Supplementary Figure 5). Impressively, during FSCV collection, we did not observe significant background drift over long recording sessions (20 minutes), that is commonly the limiting factor for long FSCV acquisitions (Hermans et al. 2008; Meunier et al. 2019; Oh et al. 2016) (Supplementary Figure 5). This indicates that the GC surface is stable under FSCV cycling *in vitro* and resists non-specific absorption of the byproducts produced by the 5-HT electrochemical reactions.

For -0.5/1.3V EW, Peak-I is less defined (Supplementary Figure 2 d-e, Supplementary Figure 3 b, c ), occurring at $0.42\pm0.05$ V, $0.50\pm0.03$ V, and $0.69\pm0.02$ V for 400, 700 and 1000 V/s scan rates, respectively. With increasing scan rate, the oxidation Peak-II shifted to the right (from $0.68\pm0.03$V at 400V/s to $0.78\pm0.02$V at 700 v/s and $1.01\pm0.01$V for 1000 V/s), while the reduction peaks shifted left-wards ($0.08\pm0.04$V , $-0.04\pm0.01$V and $-0.17\pm0.01$V  for 400, 700 and 1000 V/s, respectively). This suggests that the electrochemical kinetics of 5-HT at GC electrodes slows down with an increase in scan rates for -0.5/1.3V EW triangular waveform (Supplementary Figure 2 d, e). This is similar to what was observed for DA, with a $\Delta E$ of $0.61\pm0.03$V, $0.82\pm0.03$V and 1.18±0.01V for 400, 700 and 1000 V/s, respectively (Supplementary Figures 2 and 3). It seems, therefore, that despite the higher scan rate, the larger EW still allows GC microelectrodes to detect the secondary oxidation product of the multi-step 5-HT oxidation reaction by slowing down the overall kinetics. This confirms the influence of the EW (i.e. scan duration) and scan rate on 5-HT kinetics.

Using the modified *N*-shaped waveforms, only one oxidation peak was observed for all the scan rates (Supplementary Figure 2 f). These peaks shifted towards the right with increased scan rates (i.e., ca. 0.74, 0.90 and 1.08 V for 400, 700 and 1000V/s, Supplementary Figure 2 f). Also, the reduction peaks seem to shift towards the right, occurring at 0.17V for 1000V/s scan rate, with $\Delta E = 0.9V$, suggesting a faster kinetics compared to the one using triangular waveform with 1000V/s scan rate and same EW. However, for slower scan rates, the reduction peaks were not discriminated using this EW because they probably appear at potentials <0.2 V. It is important to note that in order to detect the positive oxidation peak (at 1.08V), we had to modify the traditional *N*-shaped waveform used for 5-HT detection at CFEs by extending the switching potential to 1.2V. (Supplementary Figure 2 f and Figure 3 a). This suggests that 5-HT kinetics at GC microelectrodes is slower than at CFE surfaces, possibly due to the difference in current densities of two dis-similar form factors of these microelectrodes.

### 3.1.2 Simultaneous Co-Detection of Dopamine and Serotonin

Once the kinetics of DA and 5-HT was investigated separately, the simultaneous co-detection of these neurochemicals was then pursued. In this case, we ruled out the use of *N*-shaped waveform and the higher scan rate of 1000 V/s that do not allow DA detection. Instead, we focused on the triangular waveform, by varying the parameters that can influence the FSCV responses, i.e. EW, holding potential, switching potential, and scan rate (Puthongkham and Venton 2020; Roberts and Sombers 2018), to obtain the best DA and 5-HT peak discrimination, considering the electrochemical kinetics at the GC microelectrodes.

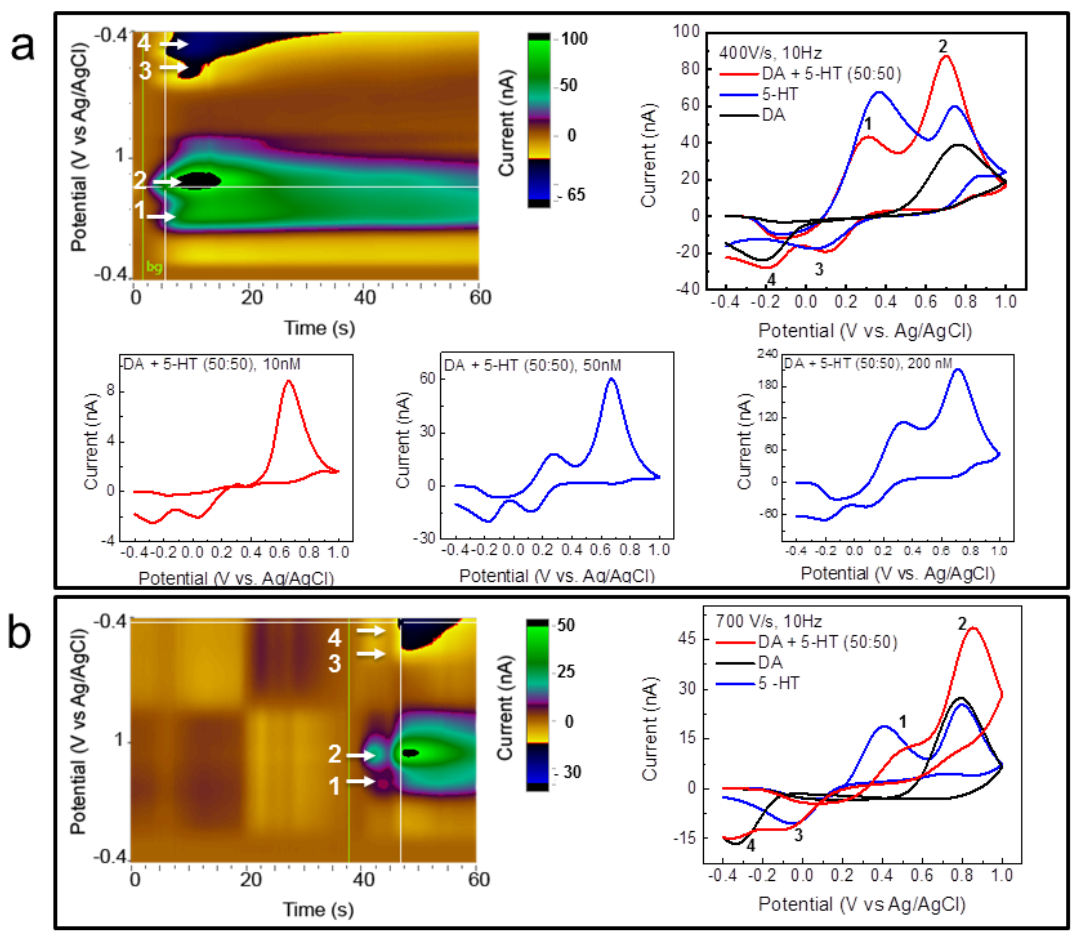

**Figure 3** Simultaneous detection of DA and 5-HT using -0.4/1 V EW at different scan rates **(a)** 400 V/s: (top left) representative color plot and (top right) background subtracted CV for 100 nM DA and 5-HT mixture (red) versus DA (black), 5-HT (blue), respectively. (bottom) Representative background subtracted CV for 10 nM, 50 nM and 200nM DA and 5-HT mixture. **(b)** 700 V/s: color plot and (left) background (right) subtracted CV for 50 nM DA and 5-HT mixture (red) versus DA (black), 5-HT (blue), respectively. Mean and standard deviations for the simultaneous

detection of different concentrations of DA and 5-HT mixtures are reported in Supplementary Figure 6. Average and standard deviation of data collected for 5 repetitions on 4 different electrodes are reported Table 3.

Under 400 V/s scan rate, -0.4/1V EW, 10Hz, 100 nM concentration of DA and 5-HT mixture, two oxidation peaks were observed (Table 3). The first oxidation was observed at 0.31±0.13V, corresponding to the 5-HT oxidation Peak-I while the second one was observed at 0.67±0.02V, slightly shifted from the 5-HT Peak-II and the DA peak (Figure 3 a top). In the representative examples reported in Figure 3 a, this peak represents 90 nM amplitude, that is the combination of the contribution from both DA and 5-HT. Additionally, two reduction peaks were clearly observed, the first at 0.09±0.02 V corresponding to the reduction peak of 5-HT and the second at -0.19±0.01V, corresponding to the DA reduction peaks. For comparison, see the separate DA (black) and 5-HT (blue) detection plots (Figure 3 a top). This marks a clear separation of the DA and 5-HT reduction peaks of around 300 mV. Thus, by having separate calibration for DA and 5-HT, it is possible to first estimate the 5-HT concentration from the 5-HT oxidation Peak-I and reduction peak, and, subsequently that of DA. Similar separations were observed for different concentration (from 10 nM to 200 nM) as shown in Figure 3 a and Supplementary Figure 6. Similar observations are made for 700 V/s scan rate and -0.4/1V EW at 10Hz (Figure 3 b). Two oxidation and two reduction peaks were observed in response to the injection of 100 nM concentration of DA and 5-HT mixture. However, while the DA and 5-HT reduction peaks were well discriminated (Table 3), the 5-HT Peak-I, at 0.50±0.04, was less defined than the one for slower scan rate (Table 3).

| 5-HT:DA | Peak-I 5-HT (V vs Ag/AgCl) | Peak-II (DA+5-HT) (V vs Ag/AgCl) | Redox Peak 5-HT (V vs Ag/AgCl) | RedoxPeak DA (V vs Ag/AgCl) |
| --- | --- | --- | --- | --- |
| 700V/s<br>-0.4/1V EW | 0.50±0.04 | 0.84±0.01 | -0.09±0.01 | -0.34±0.01 |
| 400V/s,<br>-0.4/1 V EW | 0.31±0.13 | 0.67±0.02 | 0.09±0.02 | -0.19±0.01 |

Table 3: Mean and Standard Deviation (N=3, 5 repetitions each) of the peak position of DA and 5-HT during their simultaneous detection.

Using large EW (-0.5/1.3 V), for scan rates of 400 and 700 V/s at 10Hz, it was still possible to discriminate between the reduction peaks with a separation of ~300 mV. However, it was not possible to discriminate between the oxidation peaks (Supplementary Figure 7). In this case, 5-HT showed a single oxidation peak at ca. 0.66V (400V/s) and ca. 0.87 (700 V/s), where the contribution of DA and 5-HT resulted in 60 nA amplitude. This corresponds to the summation of DA and 5-HT peak amplitudes and is in agreement with a previous study which demonstrated that extended waveform reduced the chemical selectivity of DA (Heien et al. 2003).

In summary, therefore, for -0.4/1V EW at 400 and 700 V/s scan rates, the simultaneous real-time in vitro detection of DA and 5-HT in a 50:50 mixture solution of DA and 5-HT exhibited double oxidation peaks, i.e. the characteristic Peak-I of 5-HT and a second peak, that is the summation of the DA and 5-HT (Peak-II) oxidation peaks. Further, two distinct reduction peaks were observed, each corresponding to the effects of DA and 5-HT. For larger -0.5/1.3V EW at 400 and 700 V/s scan rates it was still possible to discriminate the reduction peaks of DA and 5-HT, but not the oxidation peaks which actually converged to a single one.

### 3.1.3 Effect of Multiple FSCV

To further explore the reaction kinetics and adsorption/desorption characteristics of DA and 5-HT at the surface of GC microelectrodes, we carried out multiple FSCV (*M*-FSCV) runs (Kim et al. 2018). The adsorption of neurotransmitters from the solution on the carbon surface is well known to influence the voltammetric responses (Bath et al. 2000; Heien et al. 2003; Jang et al. 2012; Kim et al. 2018). For example, for DA detection using FSCV, holding a negative potential between voltammetric sweeps has shown to improve DA adsorption on the CFE surface, increasing the sensitivity (Bath et al. 2000; Heien et al. 2003; Jang et al. 2012; Kim et al. 2018). Furthermore, the time at which the negative constant potential is held in between scan repetitions, i.e. repetition time, influences the adsorption of the neurotransmitter on the carbon surface (Jang et al. 2012; Kim et al. 2018). Thus, considering that the adsorption behavior is specific to each analyte, depending by their intrinsic properties, the study of the adsorption characteristics of different neurotransmitters could help in (i) optimizing waveforms capable of discrimination of various analytes and (ii) mitigating fouling. Here, we used repetition time and the scan rate as parameters to study the adsorption behavior of DA and 5-HT and their differences.

First, we used a single set of *M*-FSCV scan that consisted of five consecutive triangular waveforms (cycles) with a 1 ms gap between each waveform, both for DA, 5-HT, and their mixtures. In this case, the duration of a singular waveform is 7 ms (EW: -0.4/1V at 400V/s), thus the total scan duration, considering 1ms intervals is 39 ms and the frequency is maintained to 10 Hz (61 ms at the holding potential). The five consecutive FSCV waveforms were acquired by a single *M*-FSCV scan, as shown in the color plots and corresponding background subtracted FSCV in Figure 4 a, c, 5 a, c and 6 a, c. (Other examples are reported in Supplementary Figures 8a and Figure 8 b). The redox peak amplitudes of the five subsequent FSCV showed a rapidly decreasing trend, due to the change in duration of the adsorption time (holding time in between cycles), from 61 ms to 1 ms. The adsorption properties and the kinetics of DA and 5-HT at GC microelectrodes could be determined from the rate of this decay. An example of color plot and background subtracted FSCV of a *M*-FSCV scan in response to 20 nM DA are shown in Figure 4 a, b, c. Fast decrease in peak amplitudes of the five consecutive waveforms can be observed, with a large drop of 22% noted in the oxidation peak between the first and second waveform (Figure 4 d and Table 1) and an exponential decay k = 1.37, obtained by the fitting the five consecutive DA oxidation peaks using the exponential decay function p=Ae−kt +po, where p is the peak amplitude at each consecutive cycle, A is the initial amplitude of the exponential decay, t is the cycle number, k is a positive constant term that describes the DA adsorption kinetics decrease with increasing scan number (Figure 4 d).

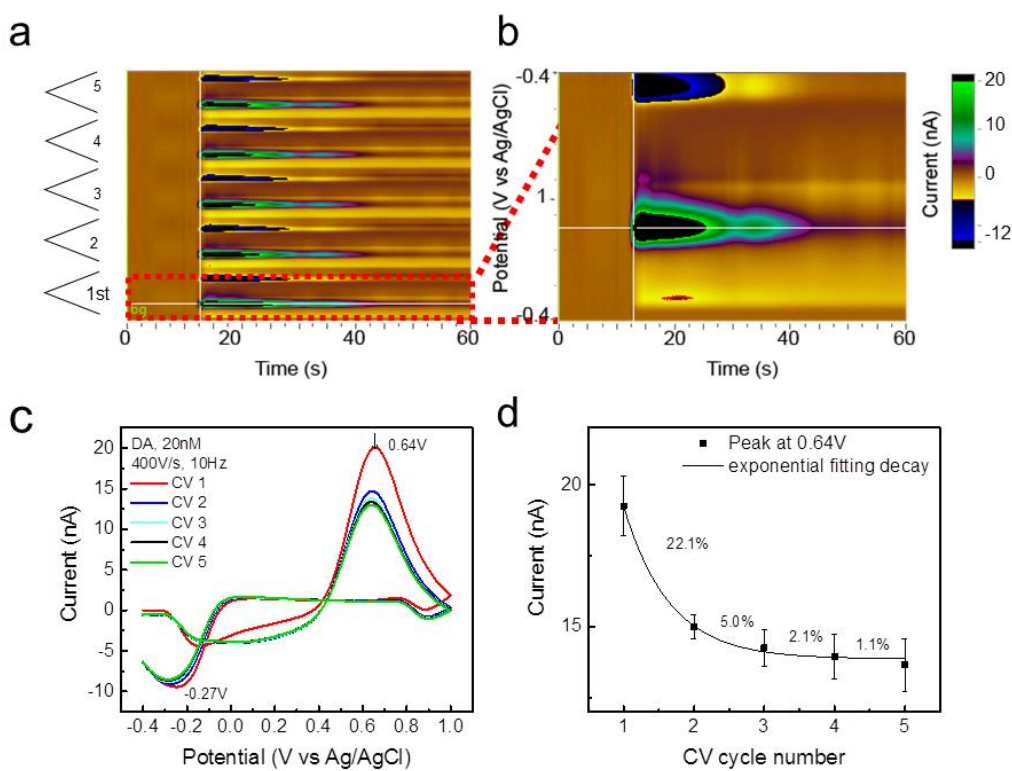

**Figure 4.** *M*-FSCV detection of DA at 400 V/sec. A single scan in *M*-FSCV consists of five consecutive triangular waveforms with a 1 ms gap between each waveform. The duration of a singular waveform is 7 ms (EW: -0.4/1V at 400V/s), thus the total scan duration, considering the 1ms intervals, is 39 ms and the frequency is maintained to 10 Hz (61 ms at the holding potential). (a-d) 20 nM DA detection: (a) representative color plot corresponding to five consecutive FSCV

waveforms acquired by a single *M*-FSCV scan with (b) magnification on the first FSCV, and (c) corresponding background subtracted CV for the five different cycle of one *M*-FSCV run. (d) percentage of the oxidation peak amplitude decay in between consecutive cycles (average and standard deviation, n=4).

The reduction peak variation is smaller, 3.8% between the first and second FSCV cycle. It is not surprising that the reduction current is not significantly influenced by the repetition time since the source of oxidation current is the DA adsorbed at the carbon surface during the holding time, while the source of reduction current is DA-o-quinone formed at the carbon surface immediately after the DA oxidation during the anodic sweep (Jang et al. 2012). Furthermore, it has been shown that DA adsorbs to carbon surface almost ten-fold stronger than DA-o-quinone (Bath et al. 2000).

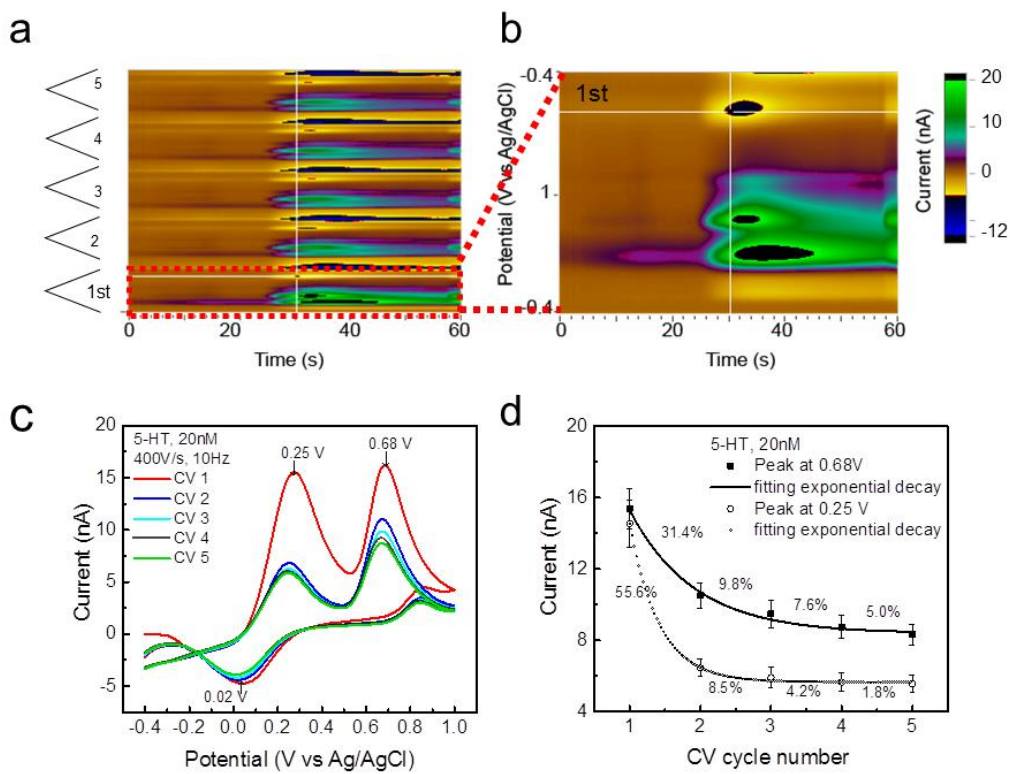

**Figure 5**. *M*-FSCV detection of 5-HT at 400 V/sec. Single set of scans in *M*-FSCV consist of five consecutive triangular waveforms with a 1ms gap between each waveform. The duration of a singular waveform is 7 ms (EW: -0.4/1V at 400V/s). Total scan duration, considering 1 ms intervals, is 39 ms and the frequency is maintained to 10 Hz (61 ms at the holding potential). (a-d) *M*-FSCV detection of 20nM 5-HT: (a) color plot corresponding to five consecutive FSCV waveforms acquired by a single *M*-FSCV scan with (b) magnification on the first FSCV, and (c) corresponding background subtracted CV for the five different cycle of one *M*-FSCV run. (d) The percentage of the oxidation peak-I and peak-II amplitudes decay in between consecutive cycles (average and standard deviation, n=4).

In the case of 5-HT (under the same conditions), a similar decreasing trend in peak amplitudes of the five consecutive waveforms were observed. However, the decrease in oxidation and reduction peak amplitudes between the first and second cycles was higher, corresponding to 31.4%, 55.6%, and 9.6% for the oxidation peaks at 0.68±0.03V (Peak-II) and 0.27±0.04 (Peak-I) and for the reduction peak, respectively (Figure 5 d and Table 4). This rapid decay in the difference of amplitude current peak with the consecutive scans (k = 1.93 for Peak-I and k = 1.01 for Peak-II) suggest that 5-HT has a stronger adsorption properties compared to DA at GC microelectrodes, similarly to what was reported for CFEs (Kim et al. 2018). Examples of color plots and background subtracted FSCVs of a *M*-FSCV scan in response to 20 nM 5-HT are reported in Figure 5 a, b, c. Other examples are reported in Supplementary Figure 8 a and Supplementary Figure 8 b.

In the case of DA and 5-HT co-detection (under the same conditions), the decay of the oxidation Peak-I (0.31±0.13V) follows the 5-HT trend, with a reduction of while the oxidation Peak-II (0.65±0.05V) seems to be influenced more by the DA behavior (k = 1.60 for Peak-I and k = 1.01 for Peak-II). The redox peaks, at -0.20±0.05V and 0.09±0.04V for DA and 5-HT respectively, are well separated and do not present high degree of decay in between consecutive cycles (Table 4). Examples of color plots and background subtracted FSCVs of a *M*-FSCV scan in response to 20nM 5-HT are reported in Figure 5 a-c. Other examples are reported in Supplementary Figure 8 a and Supplementary Figure 8 b.

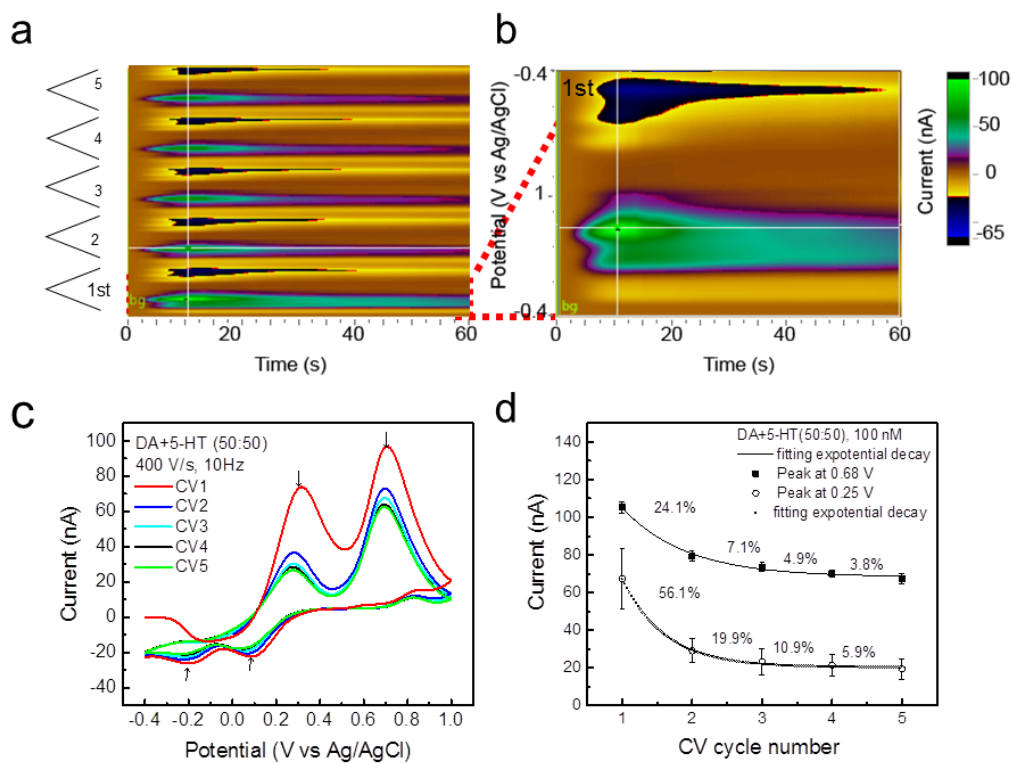

**Figure 6.** *M*-FSCV co-detection of DA and 5HT at 400 V/sec. Single scan in *M*-FSCV consists of five consecutive triangular waveforms with a 1ms gap between each waveform. The duration of a singular waveform is 7ms (EW: -0.4/1V at 400V/s). Total scan duration, considering 1ms intervals, is 39 ms and the frequency is maintained to 10 Hz (61 ms at the holding potential). (a-d) 100nM DA+5-HT (50:50) detection: (a) color plot corresponding to five consecutive FSCV waveforms acquired by a single *M*-FSCV scan with (b) magnification on the first FSCV, and (c) corresponding background subtracted CV for the five different cycle of one *M*-FSCV run. (d) percentage of the oxidation peak amplitude decay in between consecutive cycles (average and standard deviation, n=4).

.

**Table 4.** Summary of effect of *M*-FSCV on oxidation and reduction peaks of subsequent scans.

| Waveform | Scan Rate | Peak | Cycles 1-2 | Cycles 2-3 | Cycles 3-4 | Cycles 4-5 |
|---|---|---|---|---|---|---|
| **Dopamine** | | | | | | |
| Triangular | 400 v/sec | **Oxidation Peak 0.64 V** | 22.1% | 5.0% | 2.1% | 1.1% |
| | | **Reduction Peak -0.22V** | 3.8% | 1.9% | 1.5% | 0.9% |
| **Serotonin** | | | | | | |
| Triangular | 400 v/sec | **Oxidation Peak-II** | 31.4% | 9.8% | 7.6% | 5% |
| | | **Oxidation Peak-I** | 55.6% | 8.5% | 4.2% | 1.8% |
| | | **Reduction Peak** | 9.6% | 5.3% | 4.5% | 1.1% |
| | 1000 v/sec | **Oxidation Peak** | 70.4% | 18.8% | 5.3% | 4.4% |
| | | **Reduction Peak** | 31.2% | 12.5% | 8.8% | 0.9% |
| N-Shaped | 1000 v/sec | **Oxidation Peak** | 18.7% | 5.1% | ~ 0 | ~ 0 |
| | | **Reduction Peak** | ~ 0* | ~ 0 | ~ 0 | ~ 0 |
| **Dopamine + Serotonin (50:50)** | | | | | | |
| Triangular | 400V/s | **Oxidation Peak-II 0.65V** | 22.7% | 6.9% | 4.4% | 3.5% |
| | | **Oxidation Peak-I 0.3V** | 55.2% | 20.1% | 11.2% | 6.2% |
| | | **Reduction Peak (5-HT)** | 8.4% | 4.8% | 5.5% | 1.3% |
| | | **Reduction Peak (DA)** | 11.9% | 3.1% | 3.2% | 1.1% |

For 5-HT, we also used a single set of *M*-FSCV scan that consisted of ten consecutive triangular waveforms (cycles) with a 1 ms gap between each waveform. In this case, the duration of a singular waveform is 2.8 ms (EW: -0.4/1V at 1000V/s). The total scan duration, considering the 1 ms intervals, was 37 ms and the frequency was maintained to 10 Hz (63 ms at the holding potential) (Supplementary Figure 10 a and 10 b). The ten consecutive FSCV waveforms were acquired by a single *M*-FSCV scan, as shown in the color plots and corresponding background subtracted FSCV in Supplementary Figure 10 a and 10 b. The redox peak amplitudes of the ten consecutive FSCV present a more drastic decrease trend, than at lower scan rate, with a ~70% reduction of the peak amplitude between the first and the second cycle (see Supplementary Figure 10 a and 10 b and Table 4). Thus, this experiment confirmed that the use of -0.4/1V triangular waveform and higher scan rate increases the adsorption kinetic of 5-HT at the GC microelectrodes surface, and consequently, the peak amplitude decay between consecutive FSCV scan (k=1.95).

Finally, we tested the 5-HT adsorption using a single set of *M*-FSCV scan that consisted of ten consecutive N-shaped modified waveforms with a 1 ms gap between each waveform. The duration of a singular waveform is also the same here with 2.8 ms (+0.2V to -1.3V to -1V at 1000V/s). The total scan duration, considering the 1ms intervals, was 37 ms and the frequency was maintained to 10 Hz (63 ms at the holding potential). Because of the positive holding potential, the adsorption was drastically reduced (Figure Supplementary Figure 11, Table 4), with a ~18% reduction of the oxidation peak amplitude between the first and the second cycle and a very smooth decay. However, as discussed earlier, this waveform cannot be used to detect DA, and hence cannot be considered for simultaneous detection.

### 3.2 *In vivo* Co-detection of Dopamine and Serotonin

In this section, we present a proof-of-principle of *in vivo* electrochemical sensing performance of the GC microelectrodes for simultaneous FSCV detection of DA and 5-HT in the rat striatum. The procedure for recording DA and 5-HT simultaneously was adopted from the experiments by Swamy and Venton (Swamy and Venton 2007). The GC MEA was implanted in the caudate putamen where DA terminals are located for measuring the DA release evoked by electrical stimulation of the cell bodies in substantia nigra. *In vivo* evoked 5-HT concentrations are expected to be drastically lower than DA concentrations in the striatum (Stamford et al. 1990; Swamy and Venton 2007). Thus, to simultaneously detect DA and 5-HT evoked release, a pharmacologically manipulation of the 5-HT level is needed. After the MEAs were properly targeted and five consistent releases of DA were detected, carbidopa was administered 30 minutes before the injection of a synthetic precursor of 5-HT, 5-hydroxytrptophan (5-HTP), to prevent 5-HTP from being converted to 5-HT before it reaches the brain (Fuxe et al. 1971; Swamy and Venton 2007). 5-hydroxytryptophan (5-HTP) has been shown to drastically increase 5-HT release in the striatum (Fuxe et al. 1971; Swamy and Venton 2007). The color plot representations shown in Figure 7 a, b illustrate the current response following the in vivo electrical stimulation (60Hz, 250µA) at the substantia nigra region, before (a) and 45 minutes after the administration of carbidopa and 5-HT. The y-axis shows the voltage applied using a triangular waveform, starting from a starting potential of -0.4 V, ramped up to a switching potential of 1 V and back to -0.4 V holding potential. The x-axis indicates the recording time in second where the stimulation of 1 second.

Before carbidopa and 5-HTP were administered, DA current responses were recorded and a representative case of them is shown in Figure 7 a. The background subtracted voltammetry curve in Figure 7 c shows that the oxidation peak around 0.62 V and the reduction peak at ca, -0.3 V correspond to the oxidation peak and reduction peak of DA previously observed in vitro using the same waveform. 45 minutes after the 5-HTP administration (Figure 7 b, d), the color plot and background subtracted CV curve showed the characteristic 5-HT reduction peak around 0.14V, simultaneously with the DA reduction peak at ca, -0.3 V, as carbidopa does not affect the dopamine release (Fuxe et al. 1971; Swamy and Venton 2007). The oxidation peak presents a 23% current increase (Figure 7 c, e) and, in addition to the DA peak at ca. -0.3 V, the reduction peak corresponding to 5-HT around 0.14 V showed up, with a 4-times increased amplitude compared to the baseline before 5-HTP injection (Figure 7 b, d, e). We observed distorted *in vivo* CVs compared very well with the *in vitro* data. This can be attributed to the unique and dynamic chemical and biological neuro-environment in the brain with its more complex and resistive nature as compared to experiments in PBS solution in a controlled environment (Carl J. Meunier 2019; Mark DeWaele 2017; Sombers 2018). Indeed, CV peaks have been observed to shift *in vivo* due to changing electrode impedance and, more importantly, the *in vivo* reduction peaks have been observed to be smaller in magnitude (Venton and Cao 2020).

Furthermore, both from the color plot (b) and the time/current plots (Figure 7 e, f), it is possible to observe that 5-HT demonstrated a slower clearance, i.e. the oxidation peak is longer and does not come back to the baseline in the 15 second of recording session. This is likely due to the fact that pharmacological manipulation performed to elevate the basal level of 5-HT can overwhelm the capacity of serotonin uptake, thus slowing down the 5-HT re-uptake in the region. We think the results obtained here validate the proof-of-concept that 5-HT and DA can be simultaneously discriminated at GC microelectrodes using FSCV. This preliminary result is encouraging and will serve as steppingstone for further extensive *in vivo* evaluation.

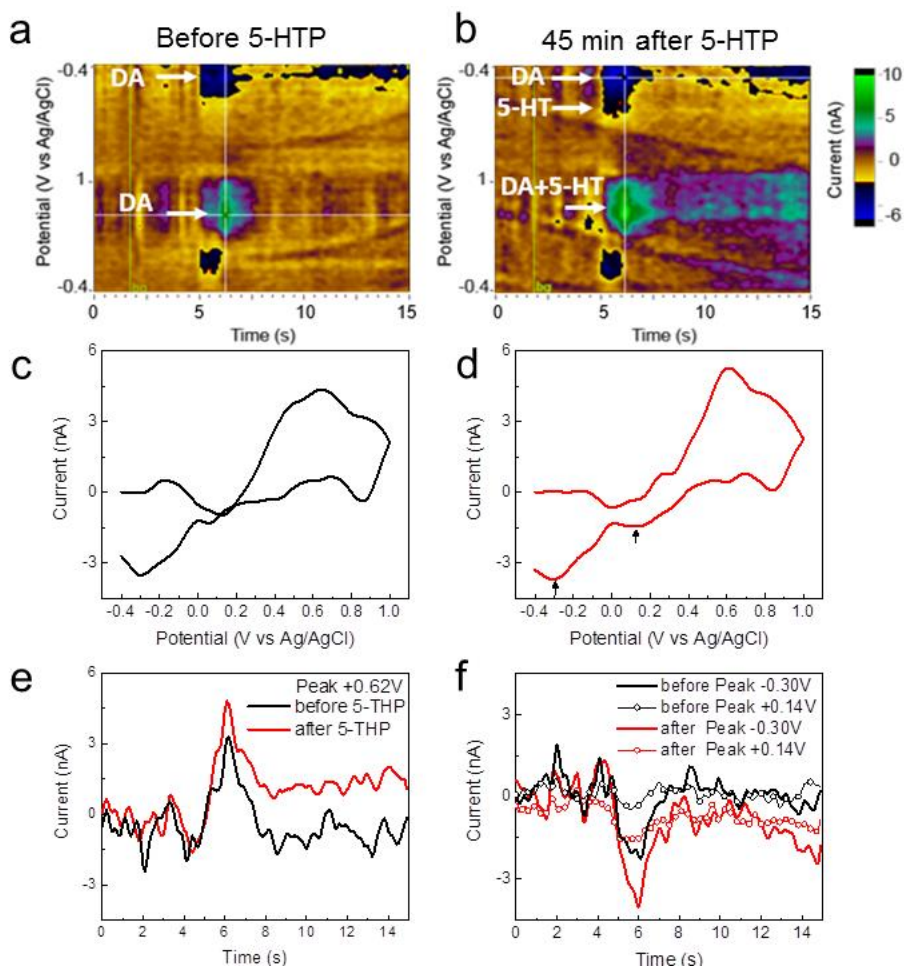

**Figure 7.** *In vivo* co-detection of dopamine and serotonin using GC microelectrodes and a triangular waveform with EW: -0.4 to 1 V, 400 V/s scan rate, recording at 10 Hz. The color plot representation of the DA (a) and DA and 5-HT (b) release by electrical stimulation (60Hz, 250μA) at substantia nigra region, before (a) and after (b) the administration of 5-hydroxytrptophan (5-HTP) to pharmacologically manipulate the 5-HT level. (a, c) The oxidation peak of DA release characteristics before 5-HTP injection was observed at 0.64 V and reduction peak at -0.3 V. b, d) 45 mins after 5-HTP injection, color plot and background subtracted FSCV presented a serotonin reduction peak around 0.2 V, together with the DA reduction peak at -0.3 V, indicating that co-detection of dopamine and serotonin can be observed in vivo using GC microelectrodes. (e, f)

comparison of the current/time plots before and after 5-HTP administration corresponding to the 0.62 V positive potential peak (e) and the 0.14 and -0.3V reduction potential peaks (f).

**3.3 What Drives Better Sensitivity of Glassy Carbon in Voltammetry?**

We had recently demonstrated that the lithography and the pyrolysis process of negative tone epoxies such as SU-8 described here produces several discontinuous basal planes and dangling carbon bonds that are rich in functional groups such as carboxyl, carbonyl, and hydroxy groups that are distributed along edges of basal planes and defects (Figure 8 a) (Hirabayashi et al. 2013). These active groups, particularly hydroxyl, carbonyl, and carboxy groups, are favorable for adsorption of cationic species such as dopamine and serotonin whose amine side chain get protonated at physiological pH.

Additionally, it was shown recently that the GC structure and atomic arrangement within the resulting graphene edges constantly evolve at different stages of the pyrolysis process, reaching maximum values for carbon pyrolyzed at around 1000 °C (Jurkiewicz et al. 2018; Sharma et al. 2018a). At this temperature, the formation of large amount of non-planar $sp^2$-hybridized carbon atoms result in the evolution of stacks of interconnected graphene fragments and curved graphene structures with well-defined protruding facets (Jurkiewicz et al. 2018; Sharma et al. 2018a). Further, it was also shown that these stacks of graphene layers have reactive edges in all directions that contain carboxyl, carbonyl, and hydroxyls functional groups.

TEM image of our electrodes is shown in Figure 8 b. In the micrograph, we can observe the presence of visible crumpled graphene-like layers produced by the pyrolysis of SU-8 at 1000 ºC (Nava et al. 2019; Woodard et al. 2018). The Raman spectra of the synthesized GC electrodes (see Figure 8c) show sharp G and D modes and broad but visible second order features (D', D'' and 2D' peaks), indicating a certain degree of graphitization in the synthesized material (see Figure 8 c) (Lee 2004). To gain better insight into the material structural properties, we performed a deconvolution of the Raman spectrum using a well-established fitting routine (see Figure 8 d; range 1100-1800 cm$^{-1}$)(Ferrari and Robertson 2000; Tai et al. 2009). From the deconvolution, we derived the values of the G peak position, the intensity ratio between the D and G modes ($I_D/I_G$), and the ratio between the slope of the linear photoluminesce background superimposed on the Raman spectrum and $I_G$. The position of the G peak (around 1605 cm$^{-1}$), higher with respect to the values typical for amorphous carbon (e.g. 1510 cm$^{-1}$), and the $I_D/I_G$ ratio, which is higher than 1, point to the presence of a substantial graphitization degree and a high sp$^2$ content in the GC material [102]. On the other hand, the low value of m/$I_G$ (close to 0 μm) indicates a low hydrogen content (Ferrari and Robertson 2000).

Both DA and 5-HT contain amine side chain, that are usually protonated at physiological pH due to their high pKa values and have net positive charge (Richter and Waddell 1983; Schindler and Bechtold 2019; TRAYNELIS 1998; Zachek et al. 2008). On the other hand, GC has n-ring aromatic system. Because this ring system allows for hybridization of electron orbitals, negative charges are concentrated above and below the atoms in the center of the ring which gives majority of the surface of GC a slight net negative charge. Further, the presence of functional groups on the outer carbon rings of GC increase its overall dipole moment. Therefore, its net negative charges coupled with large dipole moments increase the interaction of GC surfaces with dopamine or serotonin and hence GC's capability in detecting these neurochemicals even at low concentrations of 10 nM as shown in here and elsewhere (Castagnola et al. 2018; Nimbalkar et al. 2018) .

Therefore, with its dense edge planes that are rich with functional groups, GC is favorable for adsorption of cationic species and has demonstrated improved sensitivity for DA detection compared to CFEs (Puthongkham and Venton 2020; Zestos et al. 2015). Notably, defect-rich oxygen-containing carbon material surfaces have also shown an increased hydrophilicity, that help to reduce the fouling (Puthongkham and Venton 2020).

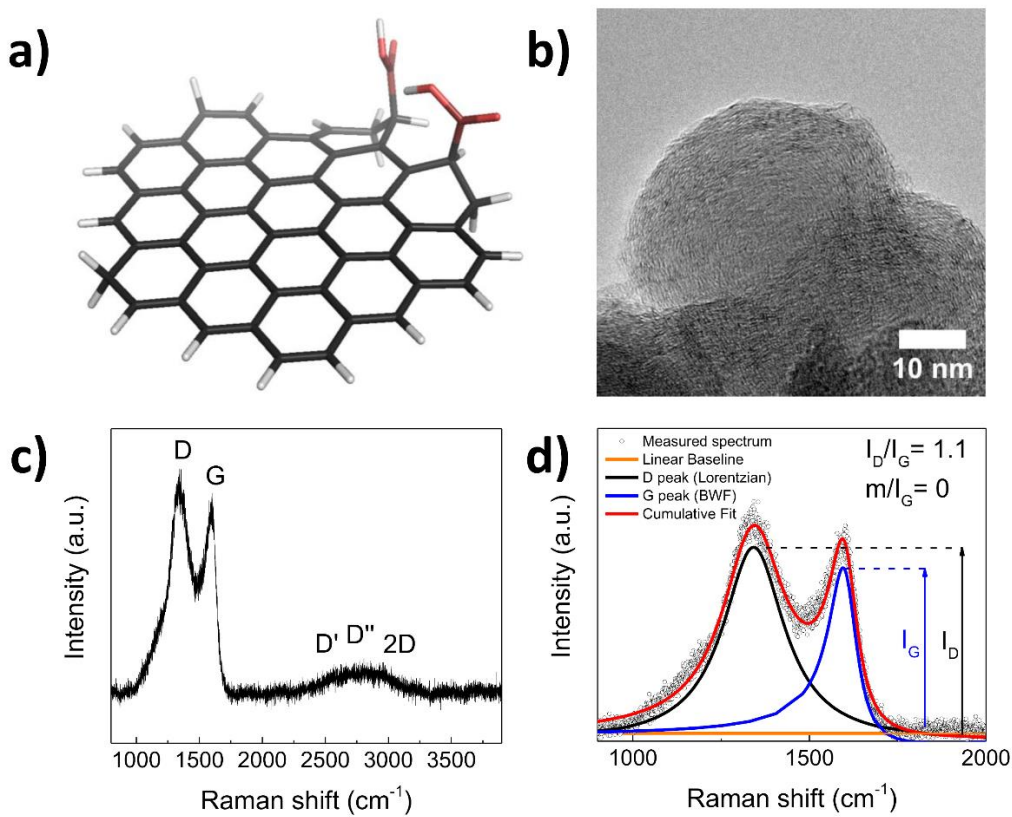

**Figure 8** (a) GC structure shown with carboxylic acid functional group(Hirabayashi et al. 2013), (b) TEM of GC showing multiple basal planes that typically end with dangling carbon bonds and functional groups. (c) Raman spectrum of a GC electrode and (d) corresponding fitting scheme (one Lorentzian peak for the D mode, one BWF peak for the G mode and a linear baseline). $I_D$ and $I_G$ are the intensities of the D and G modes respectively.

3. **Conclusions**

We present an implantable GC microelectrode array supported on polymeric substrate with four channels for *in vitro* and *in vivo* simultaneous electrochemical detection of multiple neurotransmitters, namely, dopamine and serotonin. The probe was microfabricated through the C-MEMS based pattern transfer technology recently developed by our group. These GC microelectrodes have already demonstrated higher sensitivity to dopamine and serotonin due to the numerous functional groups available on their basal planes, particularly carboxyl, carbonyl, and hydroxyl groups that are favorable for adsorption of cationic species such as dopamine whose amine side chain gets protonated at physiological pH. In this study, we focused on the characterization of the electrochemical kinetics of DA and 5-HT at GC microelectrode surfaces and gaining further insight on the adsorption/desorption mechanism of DA, 5-HT, and their combination, using multi-waveform FSCV (*M*-FSCV).

Key findings reported in this work are:

1. We demonstrate a microfabrication and validation of a glassy carbon microelectrode array that is rich with electrochemically active functional groups, good adsorption characteristics and antifouling properties.

2. We demonstrate that using optimized FSCV triangular waveform at scan rates ≤ 700 V/s and holding and switching potentials of 0.4 and 1V respectively, GC microelectrodes can simultaneously discriminate *in vitro* the reduction and oxidation peaks of serotonin and dopamine at low concentrations (10-200nM), with serotonin contributing distinct multiple oxidation peaks.

3. 5-HT oxidation involves multi-reaction steps and the background subtracted FSCV of 5-HT exhibits unique double oxidation peaks at scan rates ≤ 700 V/s and EW of -0.4V/1V. At the same EW and 1000 V/s, the two oxidation peaks seem to converge to a single peak. Using 0.5/1.3V

EW, the first oxidation peak is observed to be less defined with the 5-HT electrochemical kinetics at GC electrodes slowing down with an increase in scan rates, similarly to what was observed for DA. This confirms the influence of the EW on 5-HT kinetics.

4. There was no fouling detected on GC microelectrodes due to 5-HT after a long exposure extending over a period of 8 h.

5. *M*-FSCV results demonstrate that 5-HT, compared to DA, has a stronger adsorption property at GC microelectrodes, particularly at higher scan rates. Using -0.4/1V EW with scan rate of 400V/s, optimal for DA and 5-HT co-detection, the decay of Oxidation-Peak-I (+0.25 V) is more influence by the 5-HT trend, while Oxidation-Peak-II (+0.68V) seems influenced more by the DA behavior.

6. As a proof of principle, the GC multi-array probe was implanted in the caudate putamen of a rat brain in an acute experiment. The GC microelectrodes were able to discriminate DA and 5-HT *in vivo*.

Taken together, the results of this study demonstrate that GC multi-array microelectrodes have a compelling advantage for not only electrophysiological recording and stimulation, but also for multi-site simultaneous detection of DA and 5-HT in a stable and repeatable manner. Further, the results also demonstrate the potential of GC probes to elucidate the relationship between electrical and electrochemical signaling at synapses as part of a closed neurochemical feedback loop in the development of smart adaptive deep brain stimulation (DBS) systems.

**Supporting Information**.

Supplementary information accompanies this paper at (link)

**Competing Financial Interests:**

The authors declare no competing financial interests.

**AUTHOR INFORMATION**

Corresponding Author * Sam Kassegne • Professor of Mechanical Engineering, NanoFAB.SDSU Lab, Department of Mechanical Engineering, College of Engineering, San Diego State University, 5500 Campanile Drive, CA 92182-1323. E-mail: kassegne@sdsu.edu • Tel: (760) 402-7162.

**Present Addresses**

†Elisa Castagnola Present Addresses: Department of Bioengineering, University of Pittsburgh, Pittsburgh, Pennsylvania 15261, USA.

**Author Contributions**

**E.C.** designed and fabricated the MEA devices, performed the electrochemical detection experiments, data analysis and interpretation, and wrote major part of the manuscript. **S.T**. performed the in vivo validation and data analysis, **M.H.** helped with the data interpretation, especially with 5-HT, **G.N.** carried out TEM measurement and wrote discussion on TEM and Raman Spectroscopy. **S.N.** and **T.N.** helped in optimization and microfabrication of the MEA devices. **S.L.** and **A.O.** performed electrochemical characterization of the devices; **J.B.** wrote Matlab script for analyzing data. **C.M.** designed and supervised in vivo tests and revised the paper. **S.K.** supervised the project, structured the outline of the paper, and wrote discussion section of the paper.

**Funding Sources**

This work was supported by the National Science Foundation, the Center for Neurotechnology (CNT), a National Science Foundation Engineering Research Center [Award No. EEC-1028725].

**ACKNOWLEDGMENT**

The Authors acknowledge the support from the Center for Electron Microscopy and Microanalysis (CFAMM) at UC Riverside for TEM measurements.

**ABBREVIATIONS**

GC, glassy carbon; MEA, microelectrode array; FSCV, fast scan cyclic voltammetry; multi-waveform FSCV (*M*-FSCV); DA, dopamine; 5-HT, serotonin, EW, electrochemical window.

**Supporting Information**.

# Glassy Carbon Microelectrode Arrays Enable Voltage-Peak Separated Simultaneous Detection of Dopamine and Serotonin Using Fast Scan Cyclic Voltammetry

*Elisa Castagnola[a, e†], Sanitta Thongpang[b,c,e], Mieko Hirabayashi[a,e], Giorgio Nava[d], Surabhi Nimbalkar[a,e], Tri Nguyen[a,e], Sandra Lara [a,e], Alexis Oyawale[e], James Bunnell[e], Chet Moritz[b,e], Sam Kassegne[a,e]\**

[a] NanoFAB. SDSU Lab, Department of Mechanical Engineering, San Diego State University, San Diego, CA, USA

[b] University of Washington, Departments of Electrical & Computer Engineering, Rehabilitation Medicine, and Physiology & Biophysics, Seattle, WA

[c] Department of Biomedical Engineering, Faculty of Engineering, Mahidol University, Nakorn Pathom, Thailand

[d] Department of Mechanical Engineering, University of California, Riverside, Riverside, CA, USA,

[e] Center for Neurotechnology (CNT), Bill & Melinda Gates Center for Computer Science & Engineering; Seattle, WA 98195, USA

* corresponding author: Sam Kassegne • Professor of Mechanical Engineering, NanoFAB.SDSU Lab, Department of Mechanical Engineering, College of Engineering, San Diego State University, 5500 Campanile Drive, CA 92182-1323. E-mail: kassegne@sdsu.edu • Tel: (760) 402-7162.

**Electrochemical Characterization through Electrical Impedance Spectroscopy**

The electrochemical behavior of the microelectrodes was studied in phosphate-buffered saline solution (PBS; 0.01 M, pH 7.4; Sigma Aldrich, USA) using electrochemical impedance spectroscopy (EIS). During the EIS measurements, a sine wave (10 mV RMS amplitude) was superimposed on the open circuit potential while varying the frequency from 1 to $10^5$ Hz. EIS was carried out using a potentiostat/galvanostat (Reference 600+, Gamry Instruments, USA) connected to a three-electrode electrochemical cell with a platinum counter electrode and Ag/AgCl reference electrode. EIS plots are shown in Supplementary Figure 1. The impedance values at 10 Hz, 100 Hz and 1 kHz are 2.7E6±424.9kΩ, 494.3±58.93 kΩ, and 103.1±16.8kΩ, respectively (n = 10).

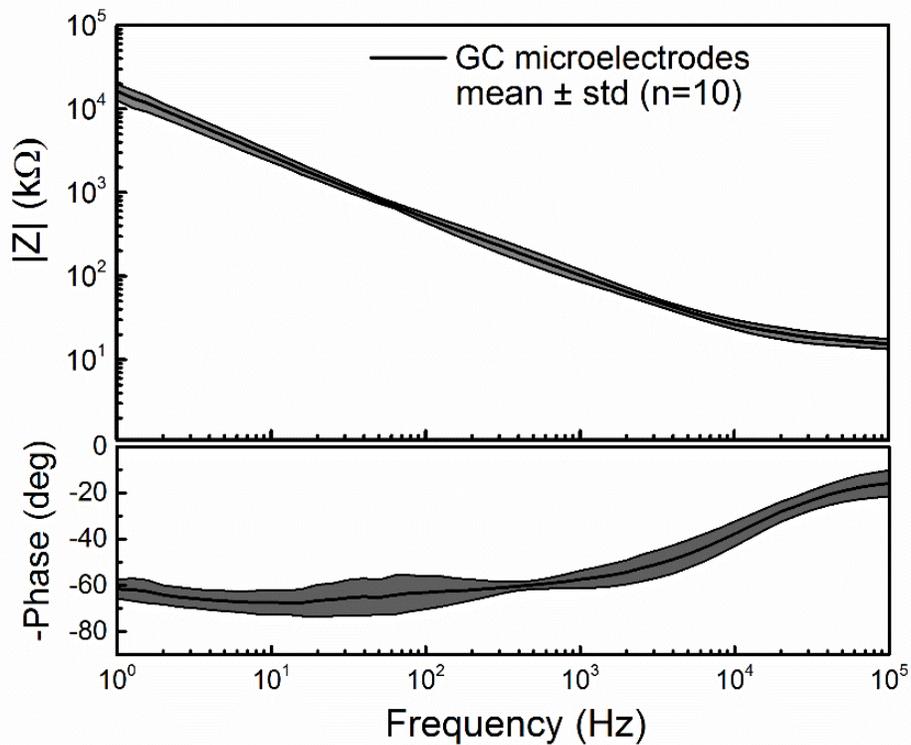

**Supplementary Figure 1** EIS results for GC microelectrodes, (mean ± std, n = 10). The impedance values at 10 Hz, 100 Hz and 1 kHz are 2.7E6±424.9kΩ, 494.3±58.93 kΩ, and 103.1±16.8kΩ, respectively (mean and standard deviation, n = 10).

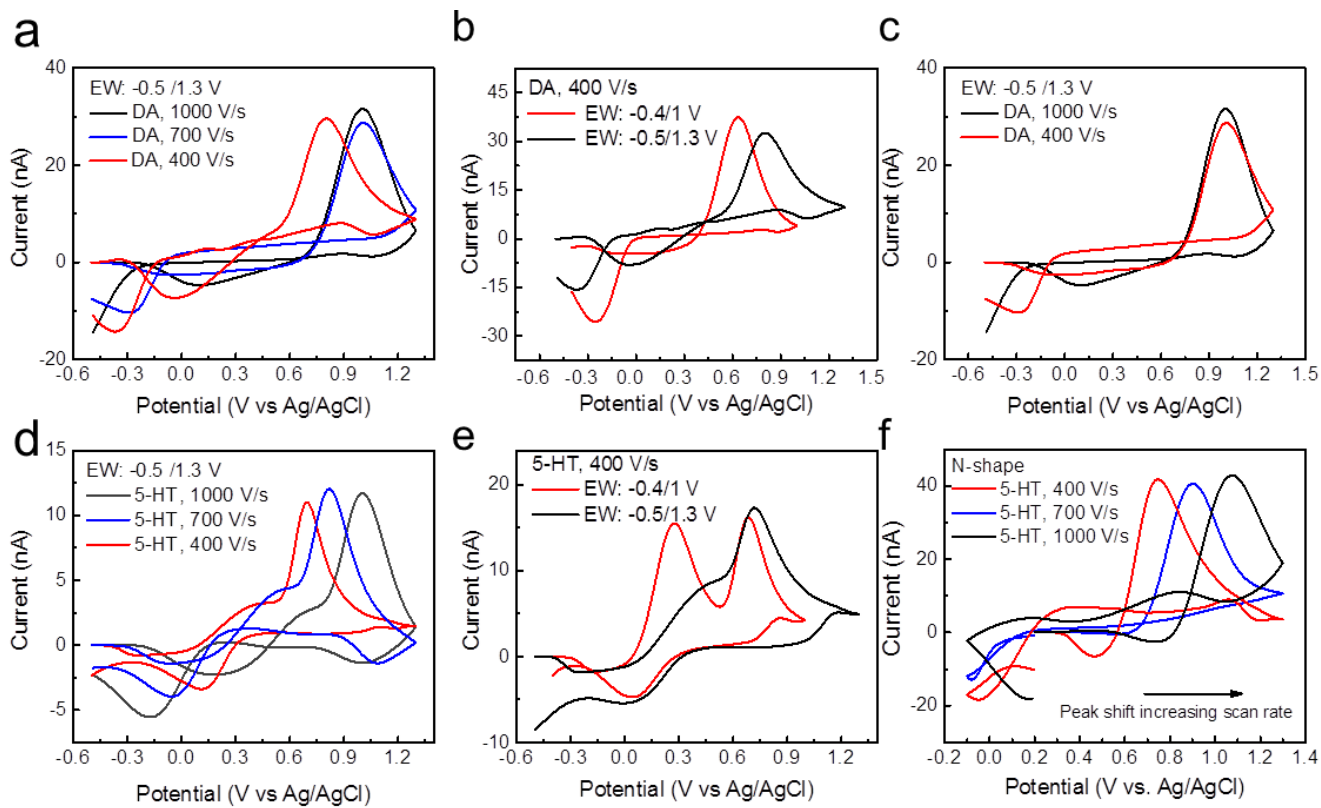

**Supplementary Figure 2 Effect of Electrochemical Window and scan rates on DA (a, b, c) and 5-HT (d, e, f) kinetics.** (a) Effect of scan rates on the response of DA. Peak shift is observed for wider EW (-0.5/1.3): for 1000 V/s, Ox = 1.01±0.02V V, Redx < -0.5V (black line), while for 700V/s, Ox = 1.02±0.05V, Redx = -0.29±0.06V (blue line), and for 400V/s, Ox = 0.79±0.01V, Redx = -0.35±0.01V (blue line), (b) Effect of EW and scan rates on the response of DA at 400V/s and (c) 1000V/s. Wider EW at 400V/s scan rate shifts Ox peaks to the right and Redox to the left. Due to the slow DA kinetics at GC electrodes using high scan rates (1000V/s), the wider EW (-0.5/1.3) allows for discrimination of Ox peak but not Redox peak. The -0.4/1V EW does not allow for DA peak discriminations. (d) Effect of scan rates on the response of 5-HT (400, 700 and 1000 V/s): at 400 V/s and -0.4/1V EW 5-HT present 2 oxidation peaks, while using wider EW (-0.5/1.3)

and scan rates ≥ 700 V/s, primary and secondary peak seem to converge. (e) Effect of scan rates on 5-HT detection using modified *N*-shaped waveform.

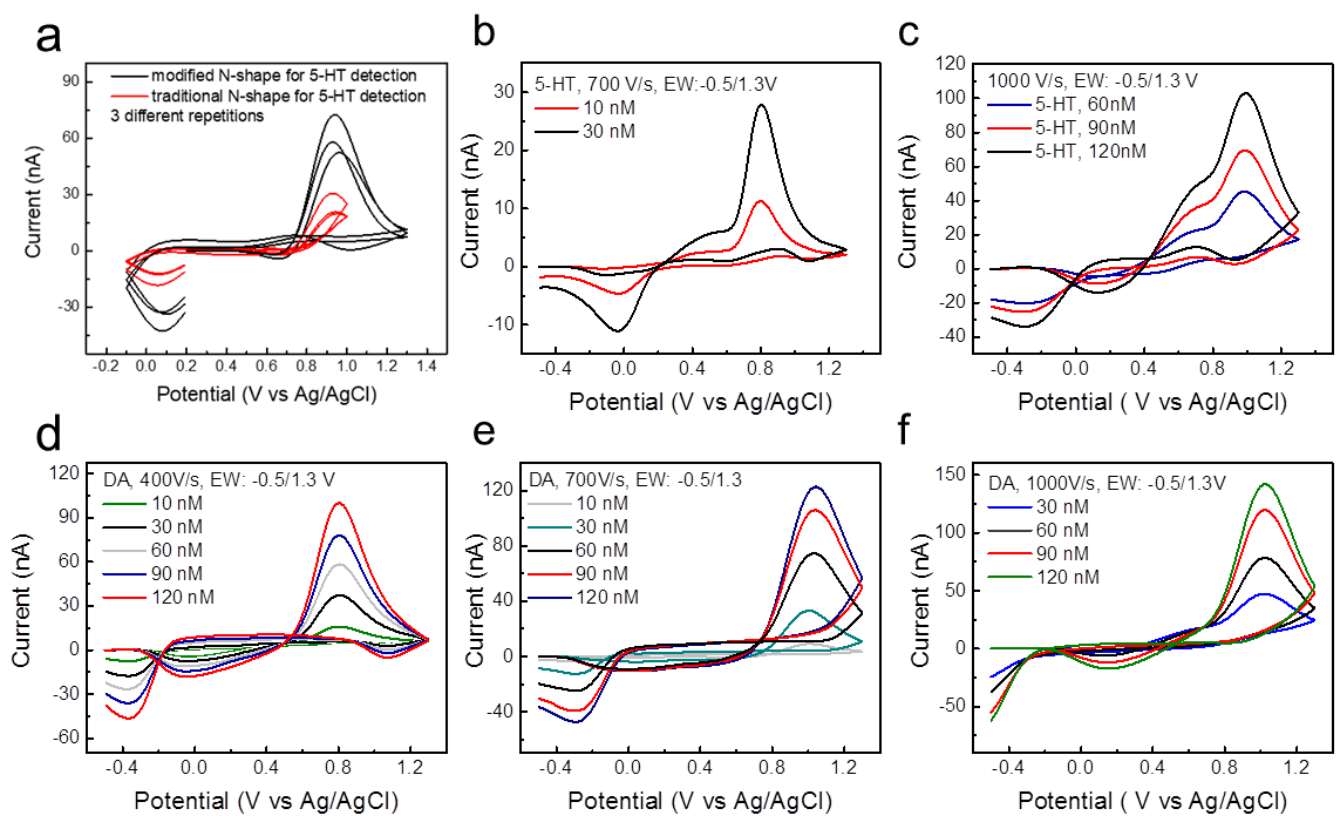

**Supplementary Figure 3** (a) Modified *N*-shape Waveform (black) versus Jackson waveform [66] (red), 3 different repetitions. It is important to note that in order to detect the positive oxidation peak (at 1.08V), we had to modify the traditional *N*-shaped waveform used for 5-TH detection at CFEs by extending the switching potential to 1.2V. (b, c) 5-HT detection (different concentrations) using the FSCV triangular waveform with -0.5 V holding and 1.3 switching potential (10 Hz), at 700V/s (b) and (c) 1000V/s. (d, e, f) DA detection (different concentrations) using the FSCV triangular waveform with -0.5 V holding and 1.3 switching potential (10 Hz), at 400V/s (d), 700 V/s (e), and 1000V/s (f).

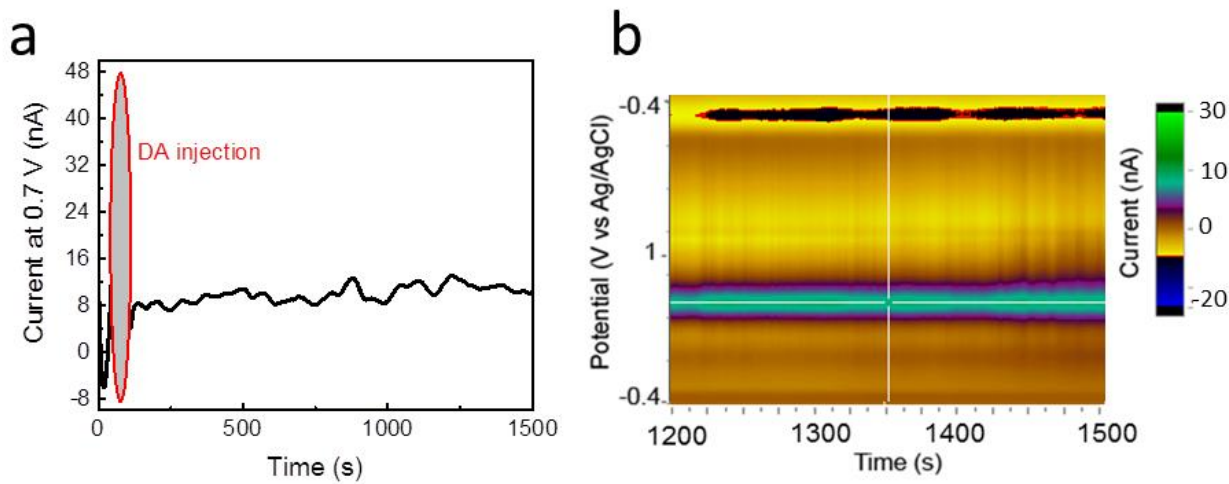

**Supplementary Figure 4 DA fouling test.** GC microelectrodes were continuously scanned in presence of 50 nM of DA using the triangular FSCV at 400 V/s. Due to our experimental set-up conditions, the potentiostat can record maximum 40 minutes of FSCV recordings at the time. (a) Representative time/current plot (corresponds to current at 0.7 V) for a recording session of 25 minutes. The current peak amplitudes in response to 50 nM of DA were stable over the entire recording session. Additionally, no significant drifting was observed in the FSCV background, demonstrating the electrochemical stability of the GC surface. (b) Example of a color plot corresponding to 5 minutes of continuous stable DA detection.

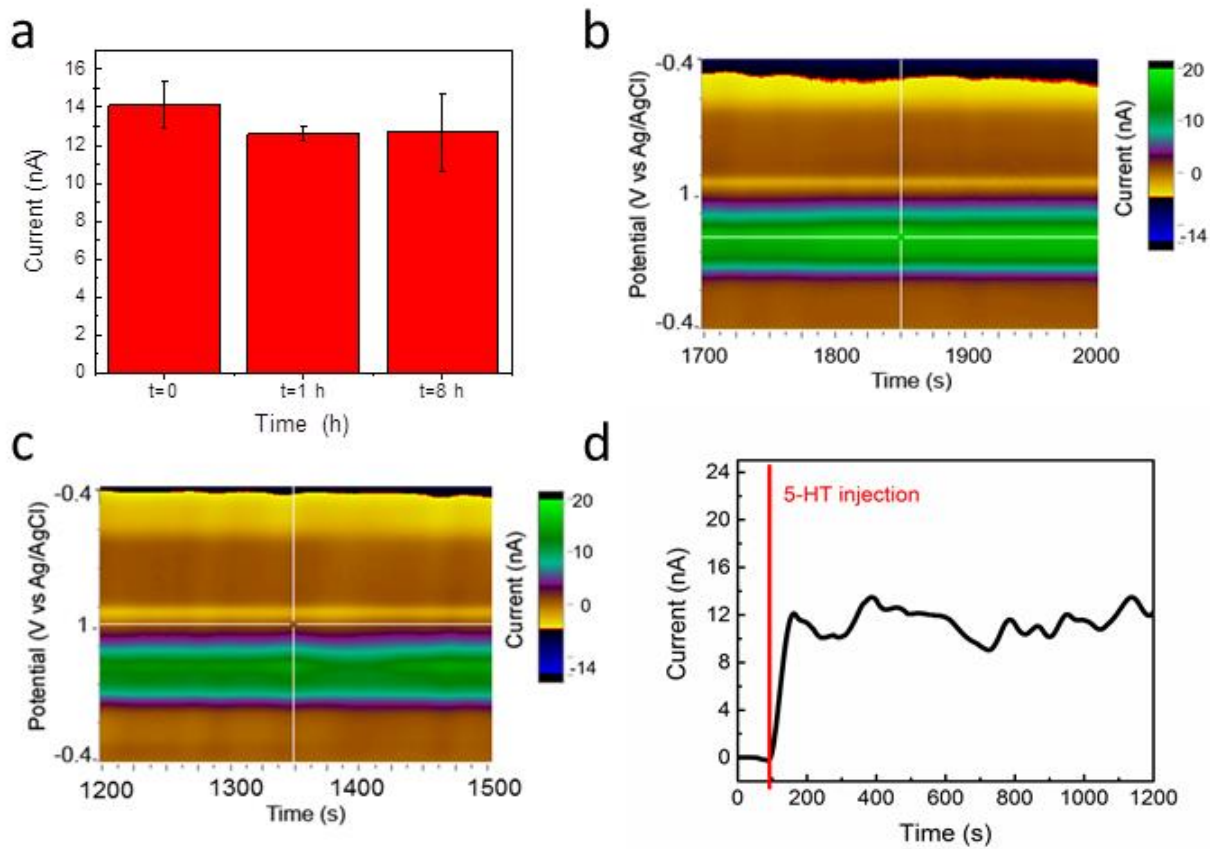

**Supplementary Figure 5: 5-HT fouling test.** GC microelectrodes were continuously scanned in presence of 50 nM of 5-HT and the 5-TH detection was monitored for 8 h using the triangular FSCV at 400 V/s. Due to our experimental set-up conditions, the potentiostat can record maximum 40 minutes of FSCV recordings at the time, thus every 20-40 minutes the PBS solution was changed and a new injection performed. (a) The current peak amplitudes in response to 50 nM of 5-HT were stable with no significant (One-way Anova, $p>0.05$) drop in detection sensitivity after 1h or 8 h. (b,c) Examples of a color plot corresponding to 5 minutes of continuous stable 5-HT detection in two different recording sessions. (d) Representative time/current plot for a recording session of 20 minutes (collected after 3 h from the beginning of the fouling test).

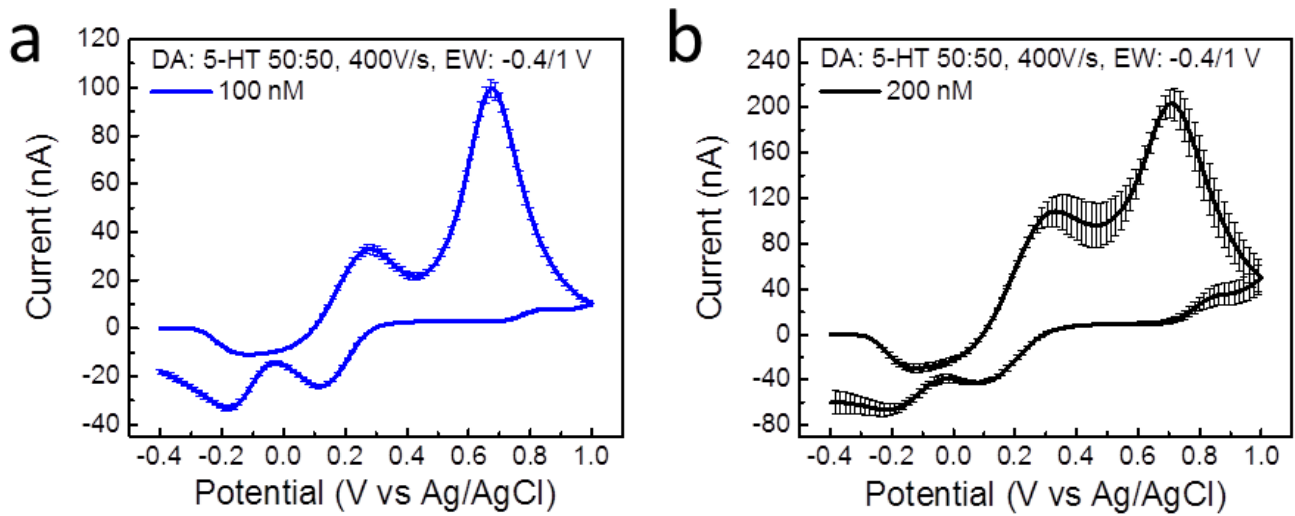

**Supplementary Figure 6** Simultaneous detection of DA and 5-HT using the FSCV triangular waveform EW: -0.4/1 V at 400 V/s: 100 nM (blue, a) and 200 nM (black, b) of DA and 5-HT mixture.

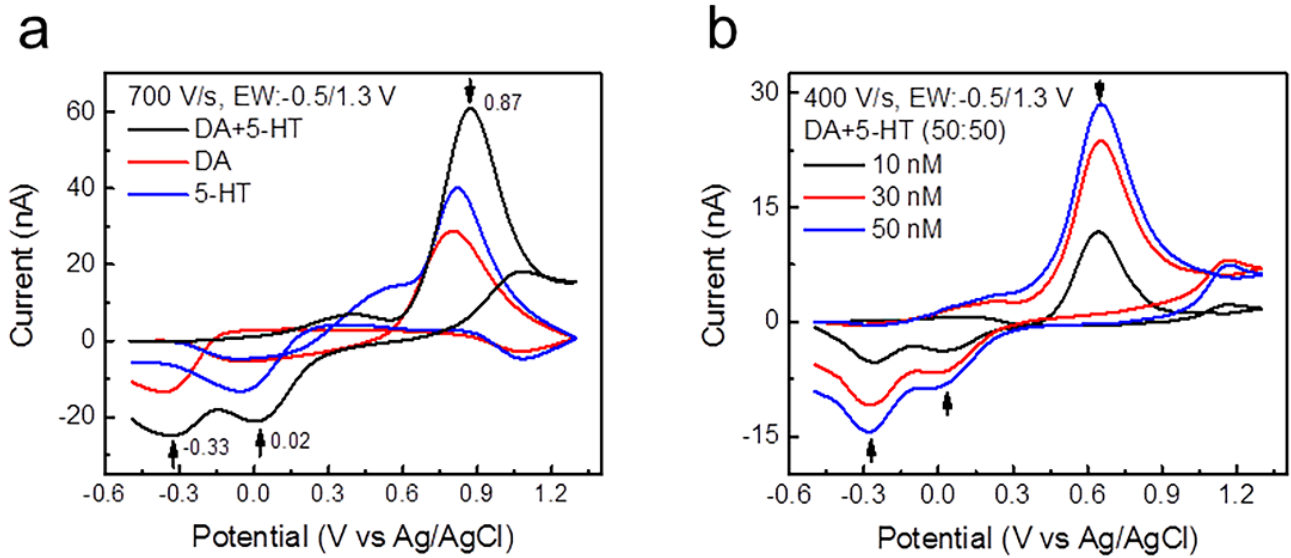

**Supplementary Figure 7** Simultaneous detection of DA and 5-HT using large EW: -0.5/1.3 V at different scan rates (a) 400 V/s: 10nM (black), 30nM (red) and 50nM (blue) of DA and 5-HT mixture, (b) 700 V/s: 100 nM DA (red), 5-HT (blue) and their mixture (black) respectively.

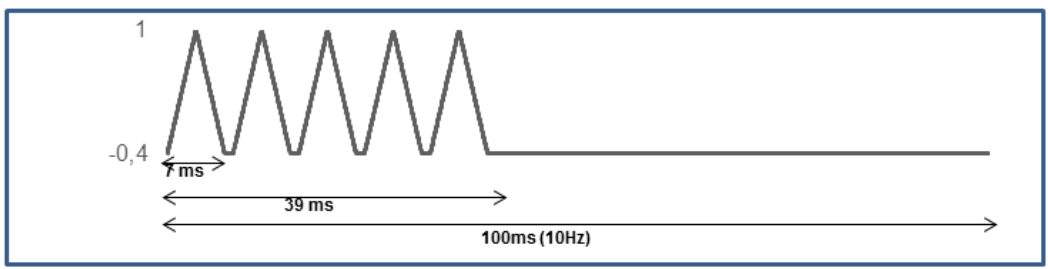

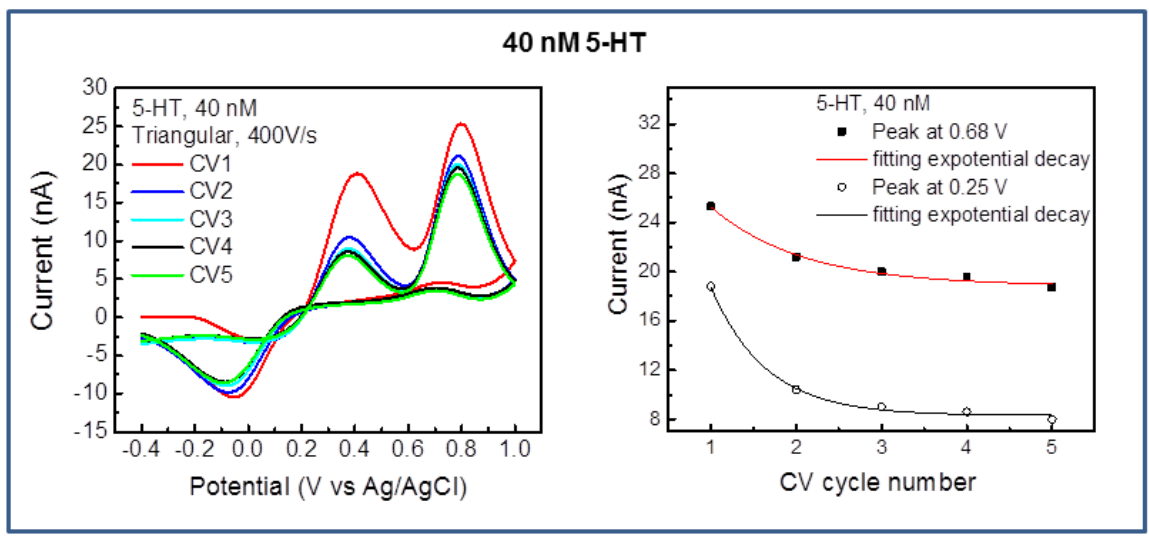

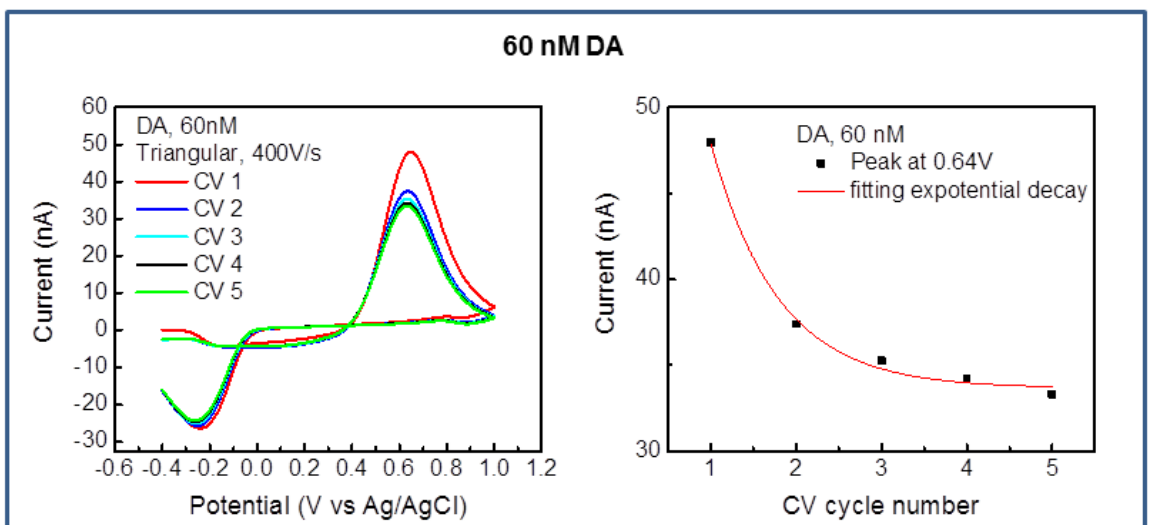

**Supplementary Figure 8 a** Other examples of oxidation peaks decay of 5-HT and DA under M-FSCV at 10 Hz **(400V/s, triangular waveform, 5 scans with 1 ms of pause, EW: -0.4/1V).**

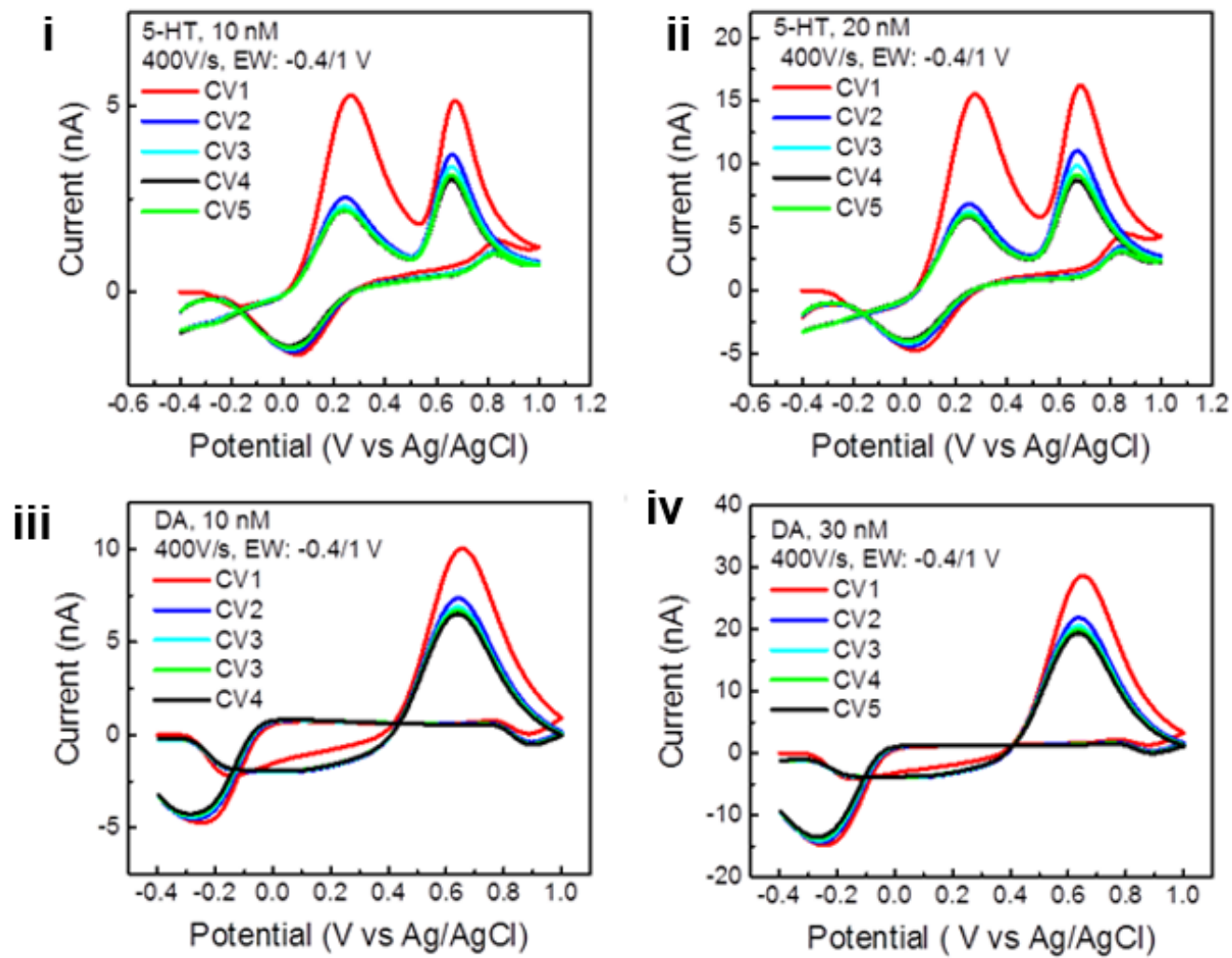

**Figure 8b.** Other examples of *M*-FSCV detection of 5-HT (i, ii) and DA (iii, iv) (different concentrations) at 10 Hz **(400V/s, triangular waveform, 5 scans with 1 ms of pause, EW: -0.4/1V).**

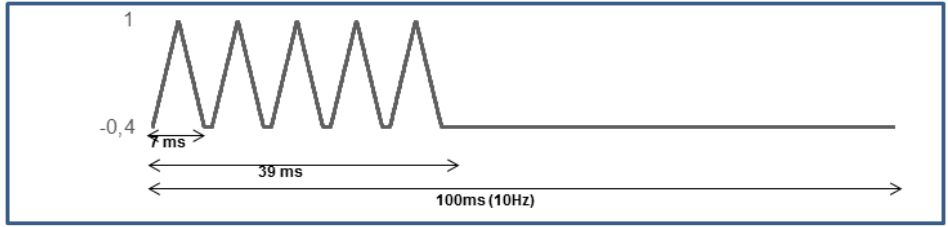

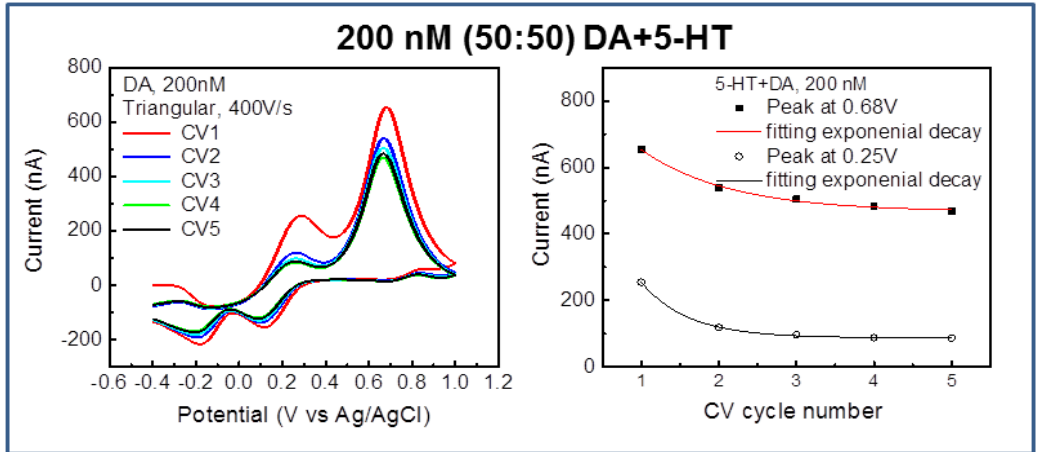

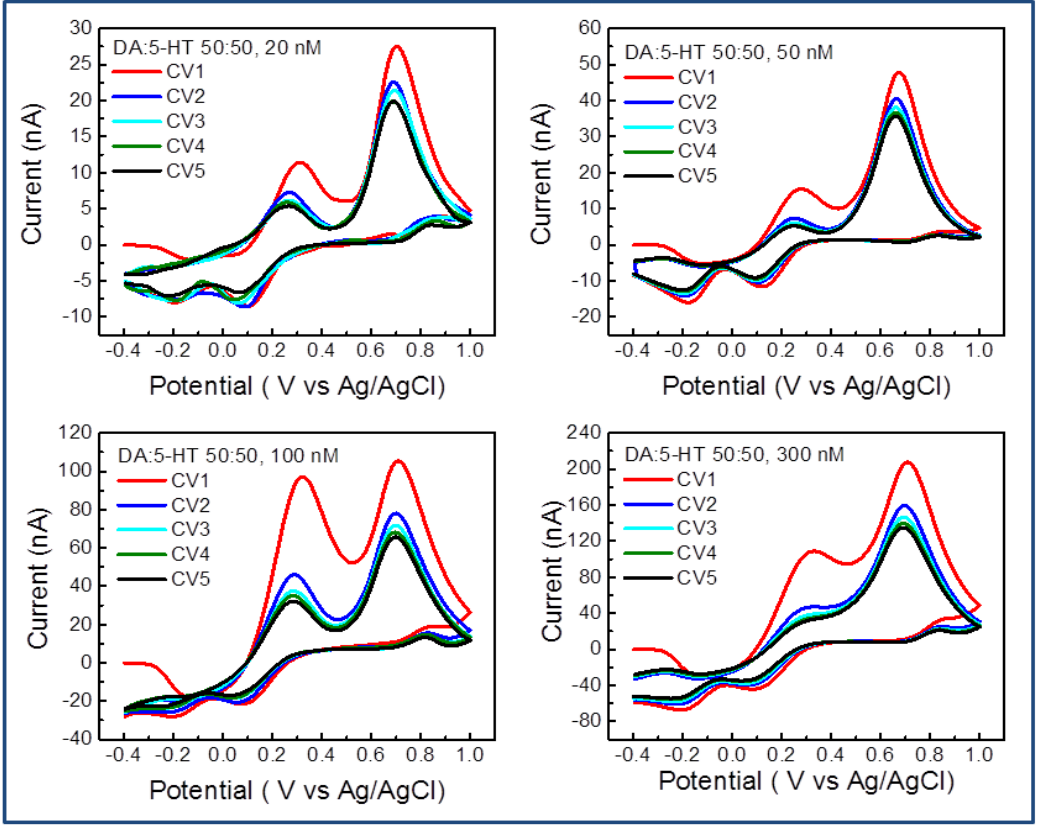

**Supplementary Figure 9** Other examples of oxidation peaks decay of mixture of 5-HT and DA (different concentrations) under M-FSCV at 10 Hz (400V/s, triangular waveform, 5 scans with 1 ms of pause, EW: -0.4/1V).

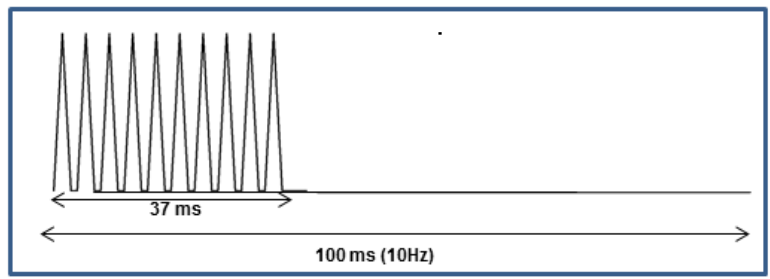

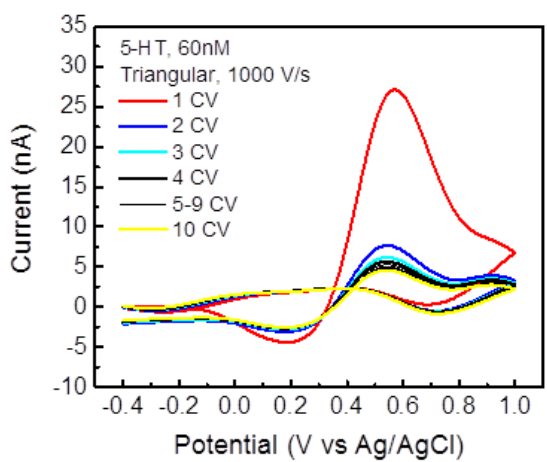

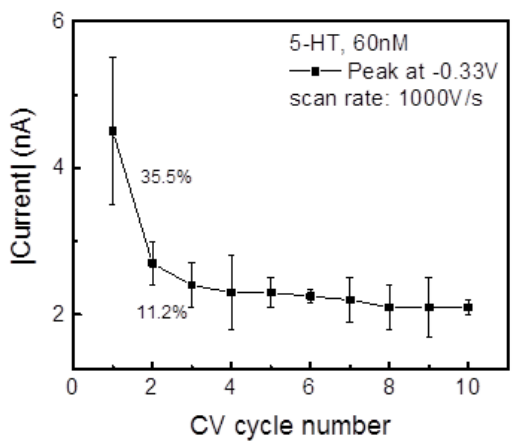
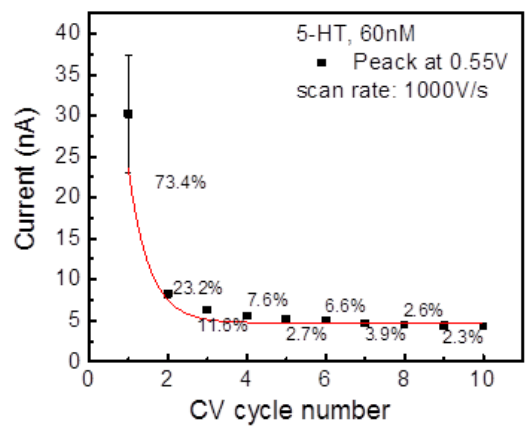

**Supplementary Figure 10 a.** Examples of oxidation and reduction peaks decay of serotonin (60 nM) under M-FSCV at 10 **Hz (1000V/s triangular waveform, 10 scans, with 1 ms of pause, EW: -0.4/1V).**

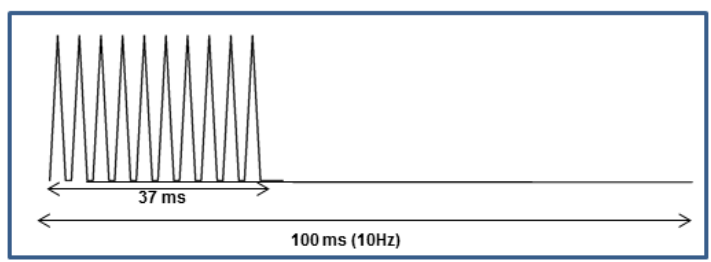

**10 nM 5-HT**

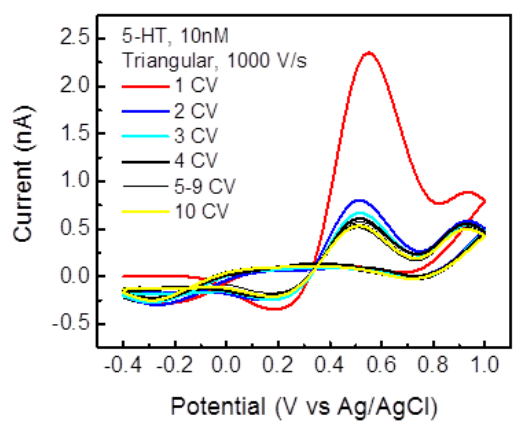

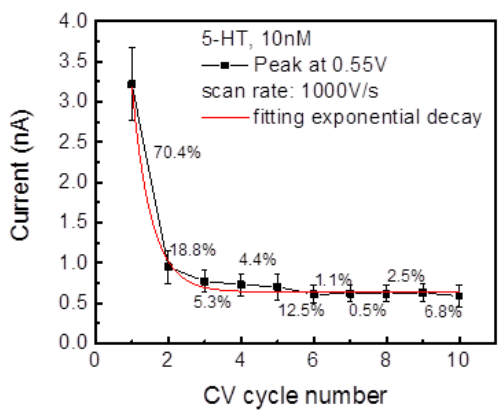
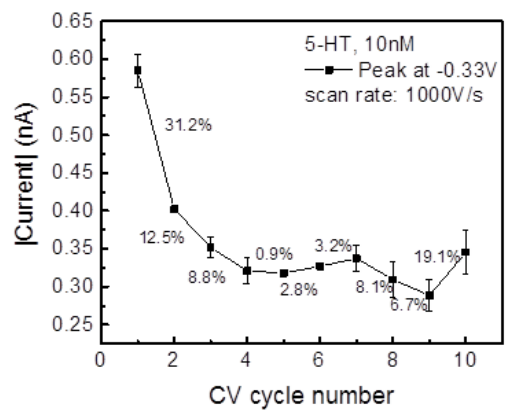

**Supplementary Figure 10 b.** Examples of oxidation and reduction peaks decay of serotonin (10 nM) under M-FSCV at 10 Hz **(1000V/s triangular waveform, 10 scans, with 1 ms of pause, EW: -0.4/1V).**

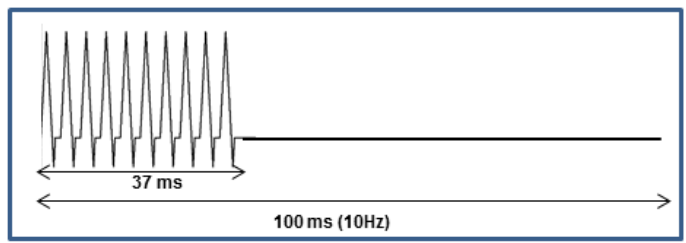

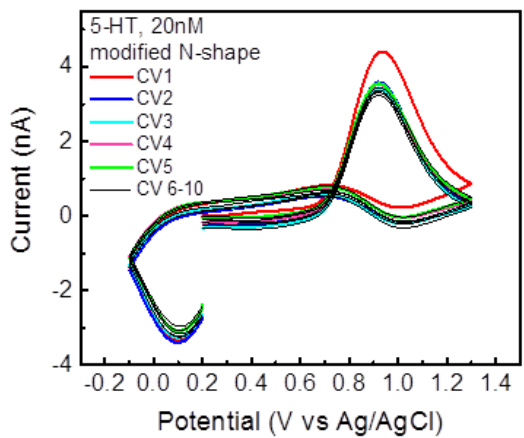

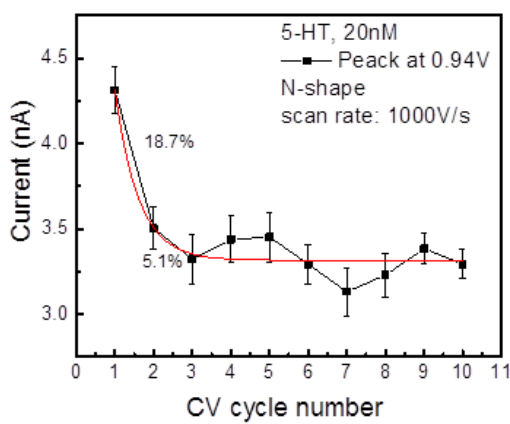
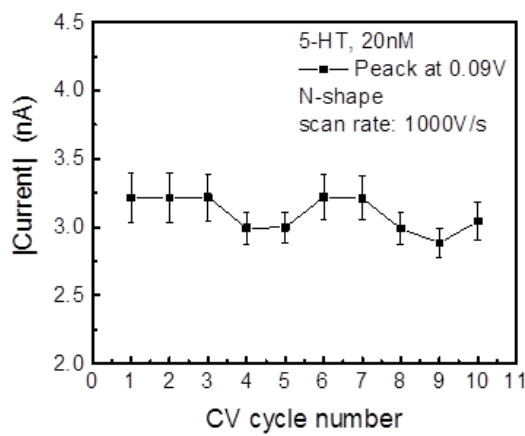

**Supplementary Figure 11.** Example of oxidation and reduction peaks decay of serotonin under *M*-FSCV at 10 Hz (1000V/s, 10 scans with 1 ms of pause, modified *N*-shape waveform).